\documentclass{aa}  
\usepackage{graphicx}
\usepackage{txfonts}
\usepackage{amsfonts}
\usepackage{amssymb}
\usepackage{booktabs}
\usepackage{xcolor}
\newcommand{\argmax}{\mathop{\mathrm{arg\:max}}}
\newcommand{\argmin}{\mathop{\mathrm{arg\:min}}}

\newcommand{\lapprox} {\, \lower3pt\hbox{$\sim$}\llap{\raise2pt\hbox{$<$}}\,}
\newcommand{\gapprox} {\, \lower3pt\hbox{$\sim$}\llap{\raise2pt\hbox{$>$}}\,}

\begin{document} 


\title{Robust construction of differential emission measure profiles \\ using a regularized maximum likelihood method}

\subtitle{}

\author{Paolo Massa
          \inst{1}
          \and
          A. Gordon Emslie
          \inst{1}
          \and
          Iain G. Hannah
          \inst{2}
          \and
          Eduard P. Kontar
          \inst{2}
          }
   \institute{Department of Physics \& Astronomy, Western Kentucky University, 1906 College Heights Blvd., Bowling Green, KY 42101, USA\\
   \email{paolo.massa@wku.edu, gordon.emslie@wku.edu}
    \and
    SUPA School of Physics \& Astronomy, University of Glasgow, Glasgow, G12 8QQ, UK\\
    }
    
\date{\today}

 
  \abstract 
   {Extreme-ultraviolet (EUV) observations provide considerable insight into evolving physical conditions in the active solar atmosphere.  For a prescribed density and temperature structure, it is straightforward to construct the corresponding differential emission measure profile $\xi(T)$, such that $\xi(T) \, dT$ is proportional to the emissivity from plasma in the temperature range $[T, T + dT]$.  Here we study the inverse problem of obtaining a valid $\xi(T)$ profile from a set of EUV spectral line intensities observed at a pixel within a solar image.}
   {Our goal is to introduce and develop a regularized maximum likelihood (RML) algorithm designed to address the mathematically ill-posed problem of constructing differential emission measure profiles from a discrete set of EUV intensities in specified wavelength bands, specifically those observed by the Atmospheric Imaging Assembly (AIA) on the NASA Solar Dynamics Observatory.}
   {The RML method combines features of maximum likelihood and regularized approaches used by other authors. It is also guaranteed to produce a positive definite differential emission measure profile.}
   {We evaluate and compare the effectiveness of the method against other published algorithms, using both simulated data generated from parametric differential emission profile forms, and AIA data from a solar eruptive event on 2010~November~3. Similarities and differences between the differential emission measure profiles and maps reconstructed by the various algorithms are discussed.}
   {The RML inversion method is mathematically rigorous, computationally efficient, and produces acceptable measures of performance in the following three key areas: fidelity to the data, accuracy in the reconstruction, and robustness in the presence of data noise. As such, it shows considerable promise for computing differential emission measure profiles from datasets of discrete spectral lines.}

\titlerunning{Construction of DEM Profiles}
\authorrunning{Massa et al.}

\maketitle


\section{Introduction}\label{sec:intro}

Extreme-ultraviolet (EUV) lines emitted by atomic species with a variety of characteristic formation temperatures provide a powerful probe of evolving physical conditions in the active solar atmosphere. However, it has not been a customary practice to use a physical source model to predict a complete set of EUV line intensities and  then directly compare these predicted intensities with those observed (the ``full forward'' approach). Neither is it feasible (for several extremely good reasons, some of which are addressed below) to use a set of observed line intensities to infer the temperature and density structure of the source (the ``full inverse'' approach). Instead, it has been a common practice in the literature to ``meet in the middle,'' using the construct of a differential emission measure (DEM) profile.
 
From the physical (``forward'') perspective, the formal definition of the (volumetric) DEM (cm$^{-3}$~K$^{-1}$) is \citep[cf. Eq. (10) of][]{1976A&A....49..239C}

\begin{equation}\label{DEM-def-physical}
\Xi \, (T) = \sum_i \, \left ( \iint_{S_{T,i}} n^2 ({\bf r}) \, \vert \nabla T \vert^{-1} \, dS_{T,i} \right)\,\,\, ,
\end{equation}
where $n({\mathbf r})$ (cm$^{-3}$) is the number density of electrons at position~${\mathbf r}$, $\nabla T$~is the gradient of the temperature distribution $T({\mathbf r})$, and the sum is over the  (one or more) constant-temperature surfaces $S_{T,i}$ within the source. For a one-dimensional source (e.g., a loop with a constant cross-sectional area) that has a monotonic temperature variation $T(s)$ along it, this reduces to the considerably simpler expression for the DEM per unit cross-sectional area:

\begin{equation}\label{DEM-def-physical-1d}
\xi(T) = n^2(s[T] \, ) \, \frac{ds}{dT} \,\,\, .
\end{equation}
This quantity, with units in cm$^{-5}$~K$^{-1}$, is the product of the square of the electron density in, and the thickness of, a layer that corresponds to a prescribed elemental range of temperatures, and it can readily be determined from a prescribed density and temperature structure.

Coming from the other (``inverse'') direction, the observed count rates (per pixel per second) from a specified pixel in an image, measured in a set of of $m$ wavelength channels, are given in terms of $\xi(T)$ by

\begin{equation}\label{eq:DEM-def}
    I_i = \int K_i (T) \, \xi(T) \, dT \, ; \qquad i = 1, \cdots, m \,\,\, .
\end{equation}
Here $K_i(T)$ (cm$^5$~pixel$^{-1}$~s$^{-1}$) is the emissivity function corresponding to the one or more spectral lines that fall within the bounds of the $i$-th wavelength channel. Here the quantity  $\xi(T) = n^2 \, ds/dT$ is the emission measure per unit area on the sky, with $s$ being the distance along the line of sight. The observational inverse problem is to deduce the form of $\xi(T)$, given a set of observed count rates $I_i$ and knowledge of the emissivity functions $K_i(T)$.

Early EUV flare observations made with the Skylab Apollo Telescope Mount (ATM) captured the emission in a set of spectral lines formed at temperatures ranging from a few $\times 10^4$~K to more than $10^6$~K, within a set of $5.5^{\prime \prime} \times 5.5^{\prime \prime}$ pixels, revealing \citep{1978SoPh...57..373E} significant temporal correlations between the emissions corresponding to different temperature ranges during several small flares. \cite{1978SoPh...57..373E} also computed the weighted DEM-related quantity ($\mu(T)$, in their notation)

\begin{equation}\label{eq:mu-T} 
\left \langle \xi(T) \right \rangle_i = \frac{\int K_i(T) \, \xi(T) \, dT}{\int K_i(T) \, dT} = \frac{I_i}{\int K_i(T) \, dT}
\end{equation}
for each observed spectral line $i$, thus allowing a relatively straightforward estimate of the evolution of the $\xi(T)$ profiles throughout the events studied. Relatively modest increases in $\langle \xi(T) \rangle$ over the durations of the various flares were found, which were interpreted as being due to a combination of a significant increase in density in the emitting volume, coupled with a steepening of the temperature gradient $dT/ds$ (e.g., from an enhanced thermal conductive flux) and hence reduced thickness of the emitting volume.

\begin{figure}[t]
\centering
\includegraphics[height=.47\textwidth, angle=90]{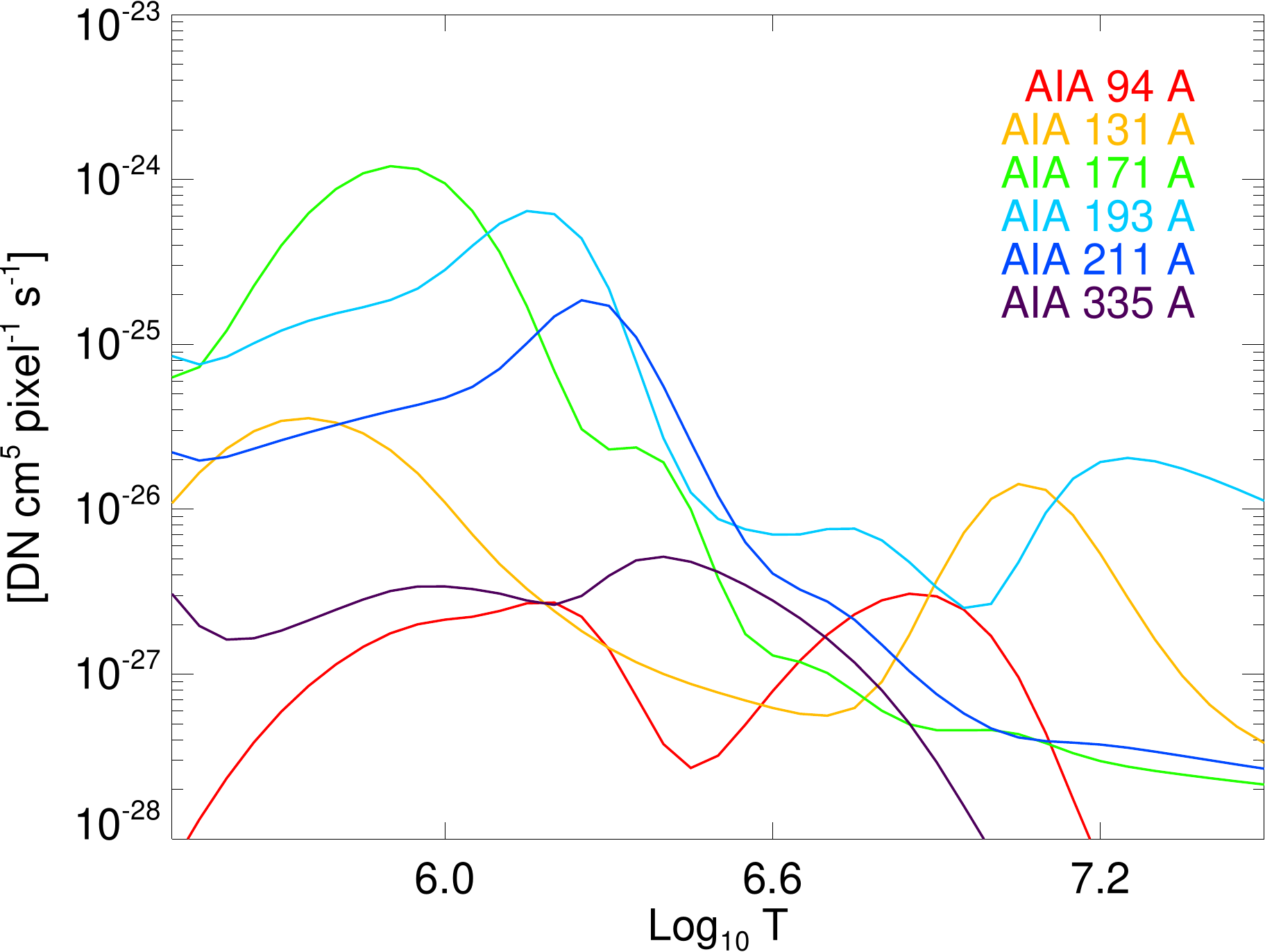}
\caption{Temperature response functions for the SDO/AIA EUV channels \citep{2012SoPh..275...41B}.}
\label{fig:aia_t_resp}
\end{figure}

Subsequent measurements by the Hinode X-Ray Telescope \citep[XRT;][]{2007SoPh..243...63G} and EUV Imaging Spectrometer \citep[EIS;][]{2007SoPh..243...19C}, and by the Solar Dynamics Observatory's Atmospheric Imaging Assembly \citep[AIA;][]{2012SoPh..275...17L} and EUV Variability Experiment \citep[EVE;][]{2012SoPh..275..115W} have provided a large amount of flare-event data from atomic species formed over a broad range of temperatures. Given the superior nature of this data, it is manifestly appropriate to go beyond the construction of average $\langle \xi(T) \rangle$ values corresponding to each wavelength channel using the prescription~\eqref{eq:mu-T}, and instead treat Eq.~\eqref{eq:DEM-def} as an integral equation to be solved for the desired source function $\xi(T)$, given the set of observed data vector values $I_i$ and the prescribed kernel functions $K_i(T)$ (shown in Fig.~\ref{fig:aia_t_resp} for the six EUV channels in the SDO/AIA instrument). 

In principle this appears straightforward: Eq.~(\ref{eq:DEM-def}) can be discretized as an $m \times n$ matrix equation

\begin{equation}\label{eq:DEM-matrix}
    I_i = \sum_{j=1}^n K_{ij} \, \xi_j \, ; \qquad i = 1, \cdots, m \,\,\, ,
\end{equation}
where the $\xi_j$ are components of a vector of $\xi(T)$ values at a set of chosen temperatures $T_j; j = 1, \cdots n$ and the $m \times n$ matrix $\mathbf{K}$ has coefficients $K_{ij}$ that represent the emissivity function for the $i$-th channel integrated over the temperature bin $[T_j, T_j + \Delta T_j]$. Naively, this has a straightforward solution in terms of the generalized inverse of the matrix $\mathbf{K}$:

\begin{equation}\label{eq:DEM-inverse-matrix}
    \xi_j = \sum_{i=1}^m ([\mathbf{K}^{T} \mathbf{K}]^{-1} \mathbf{K}^T)_{ji} \, I_i \, ; \qquad j = 1, \cdots, n \,\,\, .
\end{equation}
However, it is well understood \citep[e.g.,][]{1986ipag.book.....C} that, due to a significant inter-dependence between the rows of the matrix $\mathbf{K}$, this is an ill-posed inverse problem \citep{1963SvPhU...5..594T,1985InvPr...1..301B,1996ApJ...457..882S}. Because of the ill--conditioning of $\mathbf{K}$, the solution given by Eq.~\eqref{eq:DEM-inverse-matrix} is usually corrupted by oscillations due to noise amplification effects, and it often yields (unphysical) negative values at one or more points. Hence straightforward application of Eq.~(\ref{eq:DEM-inverse-matrix}) typically does not yield physically useful solutions. In practice, viable solutions for $\xi(T)$ must instead be found by imposing additional ``reasonableness'' requirements (e.g., smoothness, boundedness, positivity) on the solution.

Given the considerable importance of the recovered $\xi(T)$ profile in probing the physics of the emitting plasma \citep[see, e.g.,][and references therein]{2022ApJ...939...19E}, numerous authors have proposed a variety of methods for generating physically acceptable $\xi(T)$ profiles. For a general overview of such methods, we refer the reader to, for instance, Chapter~5 of \cite{2008uxss.book.....P}. Whatever the method chosen, quantitative uncertainties in the $\xi(T)$ profiles derived can straightforwardly be determined by constructing multiple realizations of the data, with each realization drawn from a range of values consistent with the level of noise in the data; synthesis of the $\xi(T)$ profiles obtained from multiple realizations of the data produces a ``confidence strip'' of $\xi(T)$ profiles \citep[cf.][]{2006ApJ...643..523B}.

Perhaps the most basic approach to solving Eq.~\eqref{eq:DEM-def} is the so-called ``EM Loci'' method, originally applied to observations of UV lines observed in stellar spectra by \cite{1987MNRAS.225..903J} and subsequently applied to solar observations from the SoHO Coronal Diagnostic Spectrometer \citep[CDS;][]{1995SoPh..162..233H} by \cite{2007ApJ...655..598C} and \cite{2007ApJ...658L.119S}, and to data from the Hinode EUV Imaging Spectrometer \citep[EIS;][]{2007SoPh..243...19C} by \cite{2009ApJ...694.1256T}. The basis of this method is to initially assume that the emitting plasma is isothermal with temperature $T_o$, so that the DEM can be approximated by a Dirac delta-function $\xi(T) = EM \, \delta(T-T_o)$, where $EM$ is the total source emission measure (cm$^{-5}$). The problem then reduces to determining the location and strength of the Dirac delta-function, i.e., the correct temperature $T_o$ and corresponding emission measure $EM$. For such a delta-function form of $\xi(T)$, Eq.~(\ref{eq:DEM-def}) reduces to $I_i = EM \, K_i(T_o)$, so that dividing an observed intensity $I_i$ by the corresponding response function value $K_i(T_o)$ gives the emission measure $EM$. Using different assumed temperatures $T_o$ yields a (generally, concave upward) set of $EM(T_o)$ points corresponding to that single observation, and repeating this for different observing channels gives a set of such ``EM Loci.'' If the source were truly isothermal, then all the EM Loci would intersect at a single point, which would then inform both the source temperature $T_o$ and the emission measure $EM$. 
However, for a nonisothermal emitting region, the various EM Loci intersect at different $(T_o, EM)$ points. Further, since the emission in a given observing channel has contributions from material at temperatures other than the assumed $T_o$, the actual value of $\xi(T)$ at any temperature is necessarily less than the value deduced from the assumption that the plasma is isothermal at that temperature. It follows that the set of local minima of the overlapping EM Loci curves forms an upper bound (``curtain'') to the $\xi(T)$ profile \citep[see, e.g., Fig.~17 of][]{2012A&A...539A.146H}. This can be useful in determining a starting point for iterative methods.

A more physical approach, that allows a priori for the nonisothermal nature of the region under observation, is to model $\xi(T)$ as a locally smooth function, using, for example, a discretized cubic spline fitting procedure. Such an approach was originally applied to data from the NRL High Resolution Telescope and Spectrograph \citep[HRTS;][]{1977ApOpt..16..870T} by \cite{1992MmSAI..63..767M}, from the Solar EUV Rocket Telescope and Spectrograph \citep[SERTS;][]{1992SoPh..137...87N} by \cite{1996ApJS..106..143B}, and from the SoHO CDS by \cite{2000A&A...363..800P}. \cite{1998ApJ...503..450K} have presented a Bayesian approach to derive $\xi(T)$ using a Metropolis Markov Chain Monte Carlo (MCMC) method, which, similar to the cubic spline fitting of the above references, applies a local smoothing, but over a scale determined by correlating the line emissivity function $K_i(T)$ \citep[$G_{u\ell} (T)$ in the notation of][]{1998ApJ...503..450K} with a parametric Gaussian-based ``Mexican Hat'' wavelet function (their Eq.~(8)). Even for simulated data generated from rather complicated forms of assumed $\xi(T)$~functions (their Figs.~7 and~8), the recovered $\xi(T)$ have a generally good overall agreement with the initially assumed forms, but unfortunately with rather large uncertainties. \cite{1998ApJ...503..450K} argue that ``a careful selection of the spectral lines used to infer the DEM is needed in order to avoid `artificially' generating structure in the DEM.''

A fundamentally different approach to recovering the DEM $\xi(T)$ involves assuming a parametric form and determining the best values of the associated parameters by forward-fitting to the observations. Such an approach, using a parametric form constructed from multiple Gaussian functions of $\log T$ (see their Eq.~(7)), was employed in the analysis of SDO/AIA data by \cite{2011ApJ...732...81A}. A similar parametric ``basis function'' approach was employed by \cite{2015ApJ...807..143C}, using a combination of finite-width Gaussians of $\log T$ (see their Eq.~(13)) and (isothermal) delta-functions. However, comparison of the $\xi(T)$ profiles derived from simulated data generated even from relatively simple functional forms of $\xi(T)$ showed on occasion spurious features at high temperatures \citep[e.g., Fig.~4 in][]{2015ApJ...807..143C}. This ``basis pursuit'' technique seeks the minimum number of nonzero basis coefficients needed to fit the data within plausible uncertainties and, to avoid (unphysical) negative solutions, it applies a positivity constraint to each of the nonzero coefficients returned. However, the underlying simplex optimization procedure sometimes fails (cf. remarks in Sect.~\ref{sec:double_gaussian}) to generate a solution that satisfies all the required constraints, leading to reconstructed DEM maps (see Sect.~\ref{sec:dem-maps} below) that contain a number of ``empty'' pixels where the method has been unable to find a solution.

The SITES (Solar Iterative Temperature Emission Solver) algorithm \citep{2019SoPh..294..135M} first normalizes the temperature response curves $K_i(T)$ for each AIA observing channel to produce a set of ``relative response curves'' that sum to unity in each of several temperature bins. Count rates obtained by forward processing of an initially assumed $\xi(T)$ profile with the normalized temperature response curves are then compared to the observed counts and the differences used to generate a correction term to $\xi(T)$, and this step incorporates a positivity constraint. This iterative process is then repeated until the $\xi(T)$ profile converges to acceptable limits. \cite{2019SoPh..294..136P} have expanded this concept into a ``Gridded SITES'' method in which pixels with similar intensities in all six EUV channels of the AIA are grouped together and a $\xi(T)$ for each group of pixels is found; this saves considerable computational time by avoiding the calculation of $\xi(T)$ profiles that are similar to those that have already been calculated for another pixel with similar data properties. Such a grouping method could obviously also be used to expedite other algorithms used to generate $\xi(T)$ profiles from datasets that comprise a set of spectral line intensities.

\cite{2010A&A...523A..44G} employed a maximum likelihood (ML; see  Sect.~\ref{sec:method-description}) approach, specifically termed a Bayesian iterative method \citep[BIM;][]{richardson1972bayesian}. They compared the results using this method with previous analyses of spectral line data from the SoHO SUMER \citep{1995SoPh..162..189W} instrument in order to validate the conclusions of \cite{2008ApJ...672..674L}. They also compared their results with the results of \cite{2010AstL...36...44S}, who had applied the BIM method (with fewer iterative steps) to spectral line data from the SPIRIT/CORONAS-F instrument. Lastly, they compared their results to broadband soft X-ray data from the Hinode XRT to validate the conclusions of \cite{2009ApJ...698..756R}, whcih were in turn based on a parametric basis function technique similar to that of \cite{2015ApJ...807..143C}, but using ``top-hat,'' rather than Gaussian, basis functions. Figs.~2 and~3 of \cite{2010A&A...523A..44G} show that the BIM method can recover rather complicated $\xi(T)$ functions (e.g., those with multiple peaks) with a reasonable degree of accuracy (we note that, in their figures, a logarithmic scale is applied to the $y$-axes). However, we show in Sect.~\ref{section:simulated_data} that ML approaches without additional regularization constraints in general produce unreliable results when applied to data consisting of only a few spectral lines, as in the case of AIA data. Another ML method proposed by \cite{1975SoPh...45..301W} and \cite{1980SoPh...67..285S} was utilized in the interpretation of Hinode XRT data by \cite{2008AnGeo..26.2999S}, and of EIS and AIA data by \cite{2014A&A...564A.130M}. Certain similarities between the performances of this ML~method and those of a genetic algorithm have been noted \citep{2008AnGeo..26.2999S}, and the use of a genetic algorithm has also proven effective \citep{2000ApJ...529.1115M} as a preconditioning step to determine the subset of spectral lines for which the corresponding~$\mathbf{K}$~matrix has minimum condition number. After this preconditioning step, the profile of $\xi(T)$ is then determined by applying the \cite{1963SvPhU...5..594T} regularization method to this selected subset of data.

Several authors have utilized regularized inversion methods; such methods impose 
a priori information on the solution in order to suppress the amplification of data uncertainties and thus obtain a a relatively stable ``smooth'' recovery of the $\xi(T)$ profile \citep[see][]{1986ipag.book.....C}. Several forms of regularized inversion, such as zeroth-order regularization, second-order regularization and maximum entropy regularization have been employed and tested using simulated EUV spectral line emission data \citep[e.g.,][]{1997ApJ...475..275J}. For example, the second-order regularized inversion method of \cite{2012A&A...539A.146H} seeks to minimize the quantity 

\begin{equation}\label{eq:regularized-equation}
\vert \vert \, \mathbf{I} - \mathbf{K} \, \mathbf{\xi} \, \vert \vert^2 + \lambda \, \vert \vert \, \xi \, \vert \vert^2 ~,
\end{equation}
where $\mathbf{I} = (I_1, \dots, I_m)$, $\mathbf{\xi} = (\xi(T_1), \dots, \xi(T_n))$, and $\lambda$ is the regularization parameter. High values of $\lambda$ smooth the solution by penalizing large deviations in $\xi(T)$, while lower values of $\lambda$ place greater emphasis on the accuracy with which the reconstructed $\xi(T)$ profile produces spectral line intensities that replicate those observed. Although these regularization approaches are generally superior to simple methods such as EM Loci, (unphysical) negative solutions can still result from over-fitting of (noisy) data; accordingly, \cite{2012A&A...539A.146H} chose the lowest value of $\lambda$ that is consistent with a nonnegative $\xi(T)$ at all temperatures $T$. The addition of the regularization term resolves many of the problems with $\xi(T)$ profiles corrupted by unphysical artifacts and generally recovers the underlying $\xi(T)$ (with uncertainties) quite well. However, analysis of simulated SDO/AIA data constructed from an assumed Gaussian $\xi(T)$ profile showed that the method can underestimate the actual $\xi(T)$ at high temperatures and overestimate it at low temperatures \citep[e.g., Figs.~11 and~13 of][]{2012A&A...539A.146H}. \cite{2013A&A...553A..10H} applied this method to a number of ``interesting'' pixels within the SDO/AIA images for a solar eruptive event on 2010~November~3, and we shall return to these results in Sect.~\ref{sec:AIA-data-application} below.

\cite{2013ApJ...771....2P} have presented a ``fast iterative regularized'' (FIR) method based on the $L^2$ norm regularization of Eq.~(\ref{eq:regularized-equation}), starting with construction of a $\xi(T)$ profile in terms of a number of prescribed basis functions $B_l(T)$, the coefficients $e_l$ of which are to be determined. \cite{2013ApJ...771....2P} also impose a positivity constraint on the recovered $\xi(T)$ values, although their six-step iterative procedure is rather involved. Their $\xi(T)$ reconstructions compare favorably with those of \cite{2012A&A...539A.146H} (their Fig.~13, noting that this comparison does not [as it possibly could] impose positivity on the reconstructions produced by the \cite{2012A&A...539A.146H} method). However, recovery of $\xi(T)$ profiles from simulated data constructed from simple Gaussian functions of $\log T$ shows (e.g., their Figs.~1 through~4) that the method tends to generate spurious features, especially at high temperatures. 

In summary, it is apparent that both maximum likelihood and regularized approaches each bring their own set of strengths (and also some weaknesses) to the overarching problem of constructing $\xi(T)$ profiles from a discrete set of optically thin spectral line intensities. Strengths include a high degree of mathematical rigor, the absence of any need to assume an a priori (parametric) mathematical form for $\xi(T)$, the ability to impose global (rather than local) smoothness constraints, the ability to authentically reconstruct model $\xi(T)$ profiles from simulated data and to determine uncertainties in the solution. Weaknesses and/or concerns include the possible generation of spurious features in the solution (particularly at high temperatures) and the occasional inability to find a solution which satisfies the imposed constraints.

We therefore here present a new ``regularized maximum likelihood'' (RML) method for the generation of $\xi(T)$ profiles from a discrete set of EUV spectral line intensities. This method combines the favorable elements of the maximum likelihood approaches of \cite{1975SoPh...45..301W}, \cite{1980SoPh...67..285S}, and \cite{2010A&A...523A..44G}, and the regularization approaches of \cite{2012A&A...539A.146H} and \cite{2013ApJ...771....2P}. We shall demonstrate, using simulated data generated from idealized single-Gaussian (Sect.~\ref{sec:single_gaussian}) and double-Gaussian (Sect.~\ref{sec:double_gaussian}) DEM forms, that this mathematically rigorous method, which does not require an a priori choice of a functional form for the solution, is nevertheless robust and always generates a (physically required) nonnegative solution. In Sect.~\ref{sec:AIA-data-application} we apply the method to the same set of representative pixels in a solar eruptive event previously studied by \cite{2013A&A...553A..10H}, and we compare the $\xi(T)$ reconstructions obtained for each of these representative pixels by several methods, noting the similarities and differences between the reconstructions obtained. Overall, we find that the RML method produces results that are generally consistent with those obtained using other methods and, unlike other methods, never generates ``outlier'' results. The method generally shows an acceptable of fidelity to the data without generating unphysical features caused by unnecessary overfitting. In Sect.~\ref{sec:summary} we summarize the results obtained and offer prospects for the implementation of this powerful DEM reconstruction method in the near-real-time prediction of solar activity.

\section{The regularized maximum likelihood (RML) method}\label{sec:method-description}

Let $g_i(x,y)$ denote the count rate per unit time detected in the $i$-th observing channel of the SDO/AIA telescope that originates within an $0.6^{\prime \prime} \times 0.6^{\prime \prime}$ pixel centered at location $(x,y)$ on the solar disk, $K_i(T)$ (cm$^5$~pixel$^{-1}$~s$^{-1}$) the temperature response function of the $i$-th AIA channel (see Fig.~\ref{fig:aia_t_resp}), and $\xi(T;x,y)$ (cm$^{-5}$~K$^{-1}$) the line-of-sight DEM at location $(x,y)$. These quantities are related by the integral equation (cf. Eq.~\eqref{eq:DEM-def})

\begin{equation}\label{eq:eq1}
g_i(x,y) \approx \int_{T} K_i(T) \, \xi(T;x,y)  \, dT ~,
\end{equation}
where the approximation is due to the fact that experimental data are inevitably corrupted by statistical (and possibly other sources of) noise. Once discretized, Eq.~\eqref{eq:eq1} becomes

\begin{equation}\label{eq:discretized}
\mathbf{g} = \mathbf{K} \, \mathbf{\xi} ~,
\end{equation}
where we have denoted $\mathbf{g}_i \equiv g_i(x,y)$, $\mathbf{K}_{ij} \equiv K_i(T_j)$, and $\mathbf{\xi}_{j} \equiv \xi(T_j;x,y)$, $i=1, \dots, m$; $j=1, \dots, n$. The essence of solving the DEM inverse problem involves finding a physically acceptable source vector $\mathbf{\xi}$ that, given an observed data vector $\mathbf{g}$, satisfies Eq.~\eqref{eq:discretized} within the bounds of observational uncertainty.

We assume that data are affected by statistical noise only, so that

\begin{equation}
\mathbf{g}_i \simeq \mathcal{P}((\mathbf{K}\mathbf{\xi})_i) ~,
\end{equation}
where $\mathcal{P}(\eta)$ denotes a Poisson random variable with a mean $\eta~>~0$. The 
maximum likelihood problem is to determine $\mathbf{\xi}^\ast$ such that

\begin{equation}\label{eq:ml-prob}
\mathbf{\xi}^\ast = \argmax_{\mathbf{\xi}\geq 0} P(\mathbf{g} \, | \, \mathbf{K} \mathbf{\xi}) ~, 
\end{equation} 
where $P$ is the likelihood function associated with a Poisson distribution, defined as

\begin{equation}\label{eq:poisson-distribution}
P(\mathbf{g} \, | \, \mathbf{K} \mathbf{\xi}) = \prod_i \frac{\exp(-(\mathbf{K} \mathbf{\xi})_i) \, (\mathbf{K} \mathbf{\xi})_i^{\mathbf{g}_i}}{\mathbf{g}_i !} ~.
\end{equation}

The condition~\eqref{eq:ml-prob} is equivalent to minimizing the negative logarithm of the likelihood function
\citep[cf. the C-statistic;][]{1979ApJ...228..939C}

\begin{equation}\label{eq:ml_eq}
\mathbf{\xi}^\ast = \argmin_{\mathbf{\xi}\geq 0} \left [ \, -\ln \, (P(\mathbf{g} \, | \, \mathbf{K} \mathbf{\xi})) \, \right ] = \argmin_{\mathbf{\xi}\geq 0} D_{\mathrm{KL}}(\mathbf{g}, \mathbf{K} \mathbf{\xi}) ~,
\end{equation}
where

\begin{equation}\label{dkl}
D_{\mathrm{KL}}(\mathbf{g}, \mathbf{K} \mathbf{\xi}) = \sum_i \, (\mathbf{K} \, \mathbf{\xi})_i - g_i \log (\mathbf{K} \, \mathbf{\xi})_i
\end{equation}
is the Kullback--Leibler divergence \citep{bertero2008iterative} and we have ignored terms in Eq.~(\ref{eq:poisson-distribution}) that depend only on the (known) quantities $\mathbf{g}_i$.

The partial derivative of $D_{\mathrm{KL}}$ with respect to each of the unknowns $\xi_k$ is

\begin{equation}\label{dkxi-gk}
\begin{split}
\frac{\partial D_{\mathrm{KL}}}{\partial \xi_k} & = \sum_i \, \left [ \, \frac{\partial}{\partial \xi_k} \sum_j K_{ij} \xi_j - \frac{g_i}{(K \xi)_i} \frac{\partial}{\partial \xi_k} \sum_j K_{ij} \xi_j \, \right ] 
= \\ 
&= \sum_i \left [ K_{ik} - \frac{g_i}{(K \xi)_i} K_{ik} \right ] = \sum_i K^{T}_{ki} \left [ {\bf{1}}_i -  \frac{g_i}{(K \mathbf{\xi})_i} \right ]~,
\end{split}
\end{equation}
where ${\bf{1}}$ is an $m$-vector with components all equal to unity, and hence the gradient

\begin{equation}\label{eq:dkxi-grad}
\nabla_\xi \, D_{\mathrm{KL}}(\xi) = \mathbf{K}^T \left( {\bf{1}} - \frac{\mathbf{g}}{\mathbf{K} \mathbf{\xi}} \right)~,
\end{equation}
where $\mathbf{g}/\mathbf{K} \xi$ is the vector with $i$-th component equal to the ratio of the respective components $\mathbf{g}_i$ and $(\mathbf{K} \mathbf{\xi})_i$. By using the Karush-Kuhn-Tucker (KKT) conditions \citep{kuhn2014nonlinear1,kuhn2014nonlinear2}, it can be shown that the minimization problem~\eqref{eq:ml_eq} is equivalent to finding a solution of

\begin{equation}\label{eq:kkt}
\mathbf{\xi} \cdot \nabla_\xi D_{\mathrm{KL}}(\xi) = 0 ~, \quad \xi \geq 0 ~,
\end{equation}
where the multiplication and the inequality between arrays are to be interpreted component--wise. This equation can be solved by means of the iterative scheme

\begin{equation}\label{eq:ml-solve}
\xi^{(l+1)} = \frac{\xi^{(l)}}{\mathbf{K}^T{\bf{1}}} \mathbf{K}^T \left(\frac{\mathbf{g}}{\mathbf{K} \xi^{(l)}}\right) \,; \qquad \xi^{(0)} = {\bf{1}}~,
\end{equation}
starting with a ``gray'' trial vector\footnote{It is not necessary to use a realistic guess (such as the base of the EM Loci plots; see Sect.~\ref{sec:intro}) as the initial estimate $\xi^{(0)}$.} with all unit values. This iterative algorithm, also known as Expectation Maximization \citep{dempster1977maximum} or, in the context of astronomical image deconvolution, Richardson--Lucy \citep{richardson1972bayesian,lucy1974iterative}, has been applied to the solution of inverse problems in several different fields, including medical imaging \citep{shepp1982maximum}, microscopy \citep{sarder2006deconvolution}, and most recently hard X-ray imaging of solar sources \citep{2013A&A...555A..61B,massa2019count,pianabook}. We refer the interested reader to \cite{bertero2008iterative}, \cite{10.1088/2053-2563/aae109} and \cite{massa2021predictive} for a comprehensive overview of the ML method and its applications.

The ML method defined by Eq.~\eqref{eq:ml-solve} is essentially the same as the BIM~method proposed by \cite{2010A&A...523A..44G} (cf. their Eq.~(22)). Also, the ML~strategy is similar to the method proposed by \cite{1975SoPh...45..301W} and \cite{1980SoPh...67..285S}, the main difference being the adoption by the latter method of empirically defined weight functions in the iterative process. Contrary to many of the methods discussed in Sect.~\ref{sec:intro}, the ML method naturally includes a positivity constraint on the solution, which prevents the generation of unphysical negative $\xi(T)$ values. However, as is evident from the simulation results below, the ML strategy is not well suited for addressing the $\xi(T)$~reconstruction problem from EUV AIA~data only, since the low number of available data channels does not permit the application of a sufficient number of constraints; this generally leads to the presence of spurious features in the reconstructions. This suggests that it would be highly advantageous to incorporate \citep{green1990use} a penalty term to allow the inclusion of a priori information on the solution. Specifically, we incorporate a term which penalizes solutions with large $\xi(T)$ values at high temperatures in order to suppress the generation of artifacts at the upper end of the temperature range under consideration. With this term added, the minimization problem~\eqref{eq:ml_eq} becomes

\begin{equation}\label{eq:rml}
\mathbf{\xi}^\ast = \argmin_{\mathbf{\xi}\geq 0} \left [ D_{\mathrm{KL}}(\mathbf{g}, \mathbf{K} \mathbf{\xi}) + \lambda \sum_j \, T_j \, \xi_j \, \Delta T_j \, \right ] ~,
\end{equation}
where $\lambda > 0$ is a regularization parameter and the penalty term represents the DEM-weighted mean temperature. The rationale for this choice of the penalty term is to select the solution, among all the ones with similar accuracy in fitting the data, that results in a $\xi(T)$ profile that is most concentrated at low temperatures. Indeed, the penalty term can be written as $(\overline{n} /3 \, k_B) \times \sum_j 3 \, n_j \, k_B \, T_j \, (ds/dT)_j \, \Delta T_j$, where $k_B$ is the Boltzmann constant. The penalty term is thus proportional to the product of the mean source density $\overline{n}$ and the quantity $\int 3 \, n \, k_B T \, ds$, representing the total thermal energy per cm$^2$ along the line of sight. The regularization constraint thus effectively seeks, among solutions that adequately fit the data, the one with the lowest thermal energy content.

For solving problem~\eqref{eq:rml}, we note that the derivative $\partial/\partial \xi_k$ of the objective function (Eq.~\eqref{dkxi-gk}) now includes an additional term $\lambda \, T_k \, \Delta T_k$, which accordingly modifies the gradient expression~\eqref{eq:dkxi-grad}. Incorporating this additional term, the iteration formula~\eqref{eq:ml-solve} becomes

\begin{equation}\label{eq:rml_solve}
\xi^{(l+1)} = \frac{\xi^{(l)}}{\mathbf{K}^T{\bf{1}} + \lambda \, \mathbf{T}} \, \mathbf{K}^T \left(\frac{\mathbf{g}}{\mathbf{K} \xi^{(l)}}\right) \,; \qquad \xi^{(0)} = {\bf{1}} ~,
\end{equation}
where $\mathbf{T} = (T_1 \, \Delta T_1, \dots, T_n \, \Delta T_n)$ is an array formed by multiplying each temperature selected for the discretization of Eq.~\eqref{eq:DEM-def} by the width of the corresponding energy bin. Since the array $\lambda \, \mathbf{T}$ has only positive entries, at each iteration the estimated $\xi^{(l)}$ profile is multiplied component-wise by a nonnegative array to yield a revised estimate $\xi^{(l+1)}$. Therefore, since the iterative scheme is initialized with an array $\xi^{(0)}$ with all positive entries, each iteration, and hence the final solution, automatically satisfies the (physically required) nonnegativity requirement. The value of the regularization parameter $\lambda$ is determined according to the \cite{morozov1966solution} discrepancy principle: starting from an intentionally high value of $\lambda$ (which leads to an over--regularized solution with a correspondingly high value of the reduced $\chi^2$ measure of fidelity to the data), we then progressively decrease $\lambda$ by a constant multiplicative factor (typically $2/3$), until we reach a value $\tilde{\lambda}$ at which the solution $\tilde{\xi}$ has a reduced $\chi^2$ lower than 1. This criterion proves to be almost always effective and leads to solutions which do not overfit the data. We term the method expressed by Eq.~\eqref{eq:rml_solve} the ``regularized maximum likelihood'' (RML) method.

We conclude this section with a few remarks.

\begin{enumerate}

\item First, in rare instances, especially in weak pixels with relatively poor statistics, it is possible that the Morozov discrepancy principle cannot be satisfied for any value of the regularization parameter $\lambda \,$; in such cases we simply adopt the solution corresponding to the initially chosen value of $\lambda$. Further, the main drawback of the progressively-decreasing-estimate approach for determining $\lambda$ is that it involves performing several reconstructions in sequence and hence substantially increases the computational cost. In a future work we will explore techniques for determining a priori the optimum value of $\lambda$, possibly exploiting the use of neural networks \citep[see, e.g.,][]{alberti2021learning}.

\item Second, we can straightforwardly apply the iterative formula~\eqref{eq:rml_solve} for each of the $N$ pixels in an AIA image in parallel by casting $\mathbf{g}$ and $\xi$ as matrices of dimension $m \times N$ and $n \times N$, respectively. We start by applying Eq.~\eqref{eq:rml_solve} to every pixel and terminate the iterative process for each pixel once that pixel satisfies the discrepancy principle constraint $\chi^2 < 1$. Pixels for which acceptable solutions have been so far obtained are then discarded when we perform the next iteration with a decreased value of $\lambda$. This approach significantly reduces the total amount of computational time required to produce acceptable $\xi(T)$ profiles for every pixel in the image. 

\item Third, similar to the ML method of \cite{2010A&A...523A..44G} and the regularized approach of \cite{2012A&A...539A.146H}, we can provide a quantitative estimate of the uncertainty of the RML~solution by means of the confidence strip approach. Specifically, we apply the procedure 25~times, each using a randomized Poisson noise perturbation of the data, and then assign the standard deviation of the recovered $\xi(T)$ values for each temperature bin as an estimate of the uncertainty of that $\xi(T)$ value. For each AIA pixel, the selection of the regularization parameter value $\lambda$ is performed just once (according to the Morozov discrepancy principle) during the reconstruction process; this value of $\lambda$ is then used for each of the 25~reconstructions, thus speeding up the uncertainty estimation process.

\item Finally, to the best of our knowledge, this is the first time that a maximum--likelihood--type algorithm has been applied to the reconstruction of $\xi(T)$ profiles from AIA data exclusively.

\end{enumerate}

\section{Results}\label{sec:results}

Throughout this section, the $\xi(T)$ profiles are generated as a function of the temperature $T$~(K). However, in order to highlight features in different temperature ranges, they are plotted as a function of the logarithm (to base $10$) of the temperature.

\subsection{Simulated data}\label{section:simulated_data}

As a test of the ability of the various inversion algorithms to accurately recover the source DEM function $\xi(T)$, several ``ground truth'' analytic forms of $\xi(T)$ were used to generate simulated SDO/AIA data in the five SDO/AIA EUV channels 94~\AA, 131~\AA, 171~\AA, 193~\AA, and 211~\AA. The 335~\AA\ EUV channel was excluded because of its relatively weak temperature response function (Fig.~\ref{fig:aia_t_resp}); it is the only AIA channel that does not dominate\footnote{The 94 \AA\ channel is never strictly dominant, but it does have a response roughly equal to that of the 131~\AA\ and 193~\AA\ channels at $\log_{10} T \simeq 6.9$ (Fig.~\ref{fig:aia_t_resp}) and so is included in the analysis.} the instrument response at any temperature (Fig.~\ref{fig:aia_t_resp}), and we have found that attempting to fit the data in this channel often introduces spurious features in the recovered solutions. This simulated data in these five channels was then inverted, using six of the algorithms described above, namely:

\begin{enumerate}
    \item the basis pursuit (BP) technique of \cite{2015ApJ...807..143C};
    \item the fast iterative regularized (FIR) methodology of \cite{2013ApJ...771....2P} ;
    \item the iterative SITES method of \cite{2019SoPh..294..135M}
    \item the regularized (REG) approach of \cite{2012A&A...539A.146H};
    \item the (un-regularized) maximum likelihood (ML) method of Eq.~\eqref{eq:ml-solve}; and
    \item the regularized maximum likelihood (RML) approach of Eq.~\eqref{eq:rml_solve} ,
\end{enumerate}
to produce recovered $\xi(T)$ forms, to be compared with the original input form as a measure of the accuracy of each method. The reconstructed $\xi(T)$ profiles for all methods involved the discretization into 12~temperature bins; this number was chosen as being more than the number of data channels (five), but not so large as to unnecessarily exacerbate the nonuniqueness aspect of the solution. The chosen temperatures (at the [logarithmic] center of each bin) are given by $\log_{10} T$~(K) = $5.85 \, (0.15 ) \, 7.50$.  For each method we calculated $\xi(T)$ profiles using 25 different realizations of the data, established by perturbing the data with Poisson noise; assessing the similarities and/or differences between the $\xi(T)$ profiles reconstructed from each data realization allows a characterization of the robustness of the method.

We find in Sect.~\ref{sec:AIA-data-application} that many AIA pixels during flares and/or solar eruptive events contain data that are consistent with EUV emission generated either by a single source containing material at a relatively small range of temperatures corresponding to quiet-Sun coronal temperatures ($\simeq 1.5 \times 10^6$~K), or by a two-temperature source with both quiet Sun and enhanced ($\simeq 10^7$~K) components. We do not dwell extensively on the various physical reasons for this (although we do offer plausible explanations for the appearance of both components), but we lean heavily on this observed two-component structure to inform the types of simulated data to be used as a test of the various DEM reconstruction methods. Specifically, the methods should be able to clearly identify and adequately characterize both low-temperature and (if they exist) high-temperature components in the DEM profile.  We therefore test the methods using simulated data generated from two main types of ``ground truth''  DEM profiles: a single Gaussian function of $\log_{10} T$ with a centroid temperature ranging from $10^{6.1}$~K -- $10^{6.4}$~K (Sect.~\ref{sec:single_gaussian}), and a double Gaussian (Sect.~\ref{sec:double_gaussian}) with both a $10^{6.1}$~K component and a higher temperature component with a centroid temperature ranging from $10^{6.6}$~K to $10^{7.2}$~K.

\begin{figure*}[ht]
\centering
\includegraphics[height=\textwidth, angle=90]{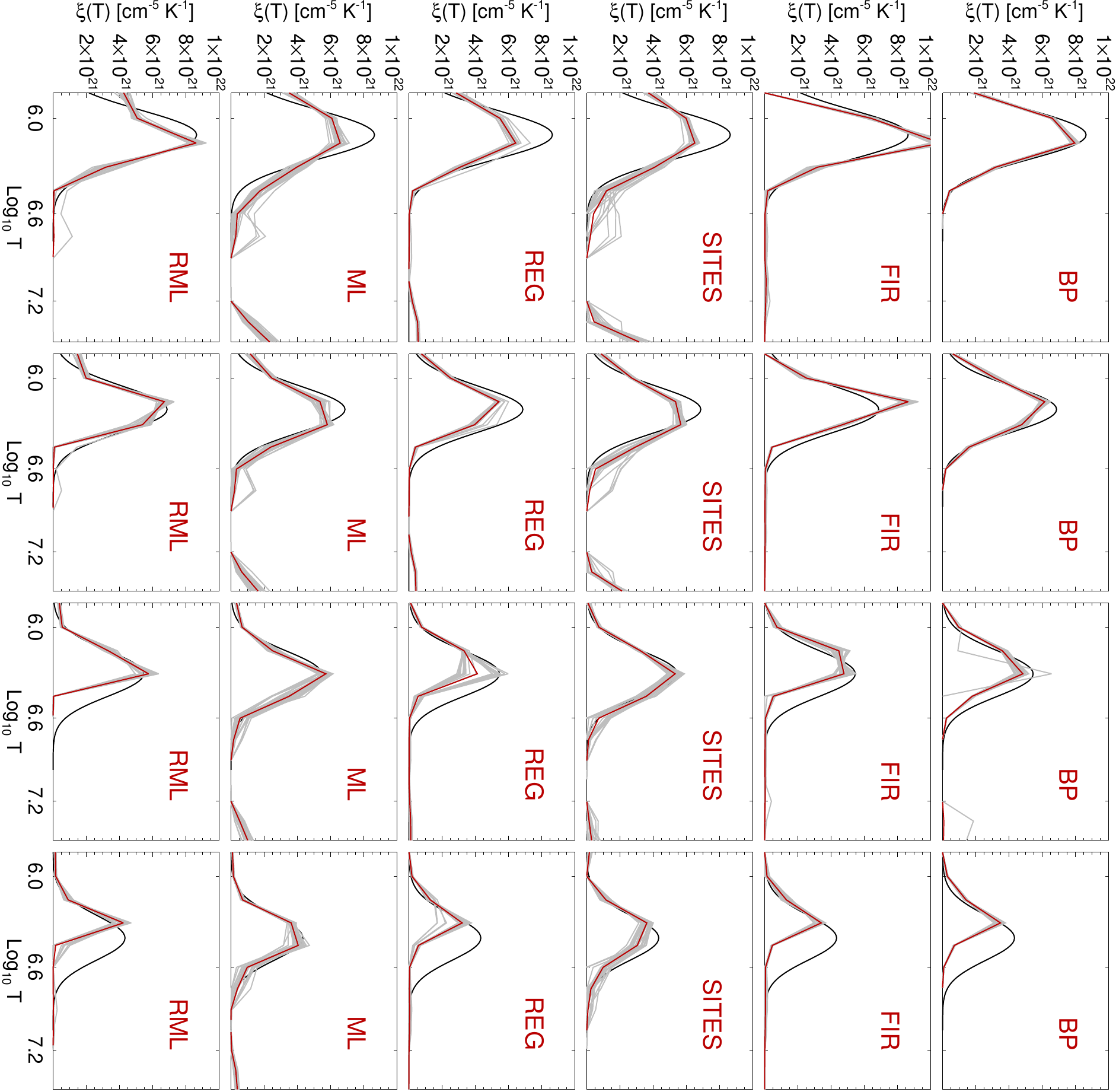}
\caption{Results of the single-Gaussian simulation tests. \textit{From top to bottom:} reconstructions of simulated Gaussian $\xi(T)$~profiles by means of the basis pursuit (BP) technique \citep{2015ApJ...807..143C}, the FIR inversion method of \cite{2013ApJ...771....2P}, the iterative SITES technique \citep{2019SoPh..294..135M}, the regularization technique of \cite{2012A&A...539A.146H} (REG), the ML method defined in Eq.~\eqref{eq:ml-solve}, and finally our proposed RML technique. The black solid lines represent the ``ground truth'' configurations, which are all Gaussian functions of $\log_{10} T$ with total Emission Measure $\text{EM} = \int \xi(T) \, dT = 10^{28}$~cm$^{-5}$ and standard deviation equal to $0.15$ dex. From left to right, the centroid temperatures are given by $\log_{10} T_c = 6.1, 6.2, 6.3,$ and~$ 6.4$. The gray lines represent the reconstructions from $25$~different realizations of Poisson-noise-perturbed data; the red lines represent the mean values of these 25~reconstructions. The $\xi(T)$ profiles are plotted as a function of the (base $10$) logarithm of the temperature~(K), so that the peak value of each Gaussian is inversely proportional to $10^x$, where $x$ is the abscissa.}
\label{fig:method-comparison-single-gaussian-linear}
\end{figure*}

\begin{table*}[ht]
\centering
\resizebox{\linewidth}{!}{\begin{tabular}{ccccccccccccc} 
\toprule
Method &\multicolumn{3}{c}{$\log_{10} T_c=6.1$} & \multicolumn{3}{c}{$\log_{10} T_c=6.2$} & \multicolumn{3}{c}{$\log_{10} T_c=6.3$} & \multicolumn{3}{c}{$\log_{10} T_c=6.4$}\\
\midrule
&Red. $\chi^2$ &\begin{tabular}{@{}c@{}}NRMSE \\ Mean\end{tabular} &\begin{tabular}{@{}c@{}}NRMSE \\ Std. Dev.\end{tabular} &Red. $\chi^2$ &\begin{tabular}{@{}c@{}}NRMSE \\ Mean\end{tabular} &\begin{tabular}{@{}c@{}}NRMSE \\ Std. Dev.\end{tabular}  &Red. $\chi^2$ &\begin{tabular}{@{}c@{}}NRMSE \\ Mean\end{tabular} &\begin{tabular}{@{}c@{}}NRMSE \\ Std. Dev.\end{tabular} &Red. $\chi^2$ &\begin{tabular}{@{}c@{}}NRMSE \\ Mean\end{tabular} &\begin{tabular}{@{}c@{}}NRMSE \\ Std. Dev.\end{tabular}\\
\cmidrule{2-13}
BP &6.7 $\pm$ 1.2 &0.10 &0.01 &7.8 $\pm$ 1.1 &0.19 &0.04 &6.1 $\pm$ 1.8 &0.48 &0.16 &6.7 $\pm$ 0.9 &1.22 &0.01 \\
FIR &2.5 $\pm$ 1.1 &0.55 &0.06 &1.8 $\pm$ 1.3 &0.54 &0.05 &2.0 $\pm$ 1.4 &0.81 &0.06 &2.5 $\pm$ 1.4 &1.29 &0.02 \\
SITES &2.7 $\pm$ 1.2 &0.70 &0.05 &2.2 $\pm$ 1.6 &0.60 &0.07 &0.9 $\pm$ 1.1 &0.24 &0.05 &0.8 $\pm$ 0.7 &0.47 &0.13 \\
REG &45 $\pm$ 14 &0.42 &0.04 &41 $\pm$ 14 &0.49 &0.06 &27 $\pm$ 23 &0.83 &0.14 &9.9 $\pm$ 15 &1.27 &0.08 \\
ML &2.7 $\pm$ 1.1 &0.61 &0.09 &2.4 $\pm$ 1.3 &0.50 &0.11 &2.2 $\pm$ 1.5 &0.42 &0.07 &1.2 $\pm$ 1.2 &0.37 &0.07 \\
RML &1.9 $\pm$ 1.5 &0.48 &0.04 &1.6 $\pm$ 1.6 &0.44 &0.05 &1.4 $\pm$ 1.5 &0.87 &0.01 &1.0 $\pm$ 0.6 &1.41 &0.09 \\
\bottomrule
\end{tabular}}
\caption{Metrics for the single Gaussian ``ground truth'' model test. The rows represent the six different reconstruction methods used (see text), while the columns are labeled by the (logarithm of the) centroid temperature of the Gaussian. For each case, the column ``Red.~$\chi^2$'' is the mean $\pm$ standard deviation of the reduced $\chi^2$ values over the 25 different realizations of the simulated data, while the next two columns measure the mean of the normalized root mean square error between the 25 recovered $\xi(T)$ profiles (a measure of the accuracy of the recovered profile) and its standard deviation (a measure of the robustness of the recovered profile to the introduction of data noise).}
\label{tab:metrics_single_gauss}
\end{table*}

\subsubsection{Single Gaussian forms}\label{sec:single_gaussian}

Fig.~\ref{fig:method-comparison-single-gaussian-linear} shows the results using simulated ``ground truth'' $\xi(T)$ that take the form of a Gaussian function of $\log_{10} T$. For each method we show the results for 25 different realizations of the data, established by perturbing the data with Poisson noise; the red line is the mean of these 25 reconstructions, while the gray lines show the individual reconstructions for each realization. From left to right, the various columns represent ``ground truth'' $\xi(T)$ forms with a total\footnote{The ``ground truth'' $\xi(T)$ profiles are Gaussians with different centroid temperatures but the same standard deviation in $\log_{10} T$. Thus the total Emission Measure $\text{EM} = \!\! \int \xi(T) \, dT \!\! = \!\! \int \xi(T) \, T \, d \ln T = (\ln 10) \int \xi(T) \, T \, d \log_{10} T$ scales as the peak value of the Gaussian times its centroid temperature $T_c$. It follows that, in order for all the ``ground truth'' profiles to have the same EM, the peak value of each profile must be inversely proportional to its centroid temperature $T_c$ = $10^{\log_{10} T_c}$. This pattern is evident in Figure~\ref{fig:method-comparison-single-gaussian-linear}.} Emission Measure $\text{EM} \!\!=\!\! \int \xi(T) \, dT = 10^{28}$~cm$^{-5}$ and a standard deviation equal to 0.15~dex; the centroid temperatures are given by $\log_{10} T_c \!=\! 6.1, 6.2, 6.3,$ and~$6.4$.

Table~\ref{tab:metrics_single_gauss} shows the values of two metrics that collectively measure the fidelity, accuracy, and robustness of each DEM reconstruction method. The first is the reduced $\chi^2$ metric, which measures the 
fidelity of the method with respect to the data: we computed the sum of the squared differences (normalized by the corresponding squared uncertainty) between the original five-channel SDO/AIA spectral line data and the set of line intensities produced by substituting the recovered $\xi(T)$ profiles into Eq.~\eqref{eq:DEM-def}; this quantity is then normalized by the number of degrees of freedom. The second metric is the ``normalized root mean square error'' (NRMSE), the root--mean--squared difference between the ``ground truth'' $\xi_j^{\text{GT}}$ and reconstructed $\xi_j^{\text{REC}}$ profiles, normalized by the mean intensity of the ``ground truth'' profile: 

\begin{equation}
\text{NRMSE} = \frac{\sqrt{\frac{1}{n} \sum_{j=1}^n \left( \xi_j^{\text{GT}} - \xi_j^{\text{REC}} \right)^2}}{\frac{1}{n} \sum_{j=1}^n \xi_j^{\text{GT}}} ~.   
\end{equation}
The mean value of this metric over the 25 different noisy data realizations provides a measure of the accuracy of the reconstruction, while its standard deviation over this set of realizations is a measure of the robustness of the method, i.e., the sensitivity of the recovered $\xi(T)$ profile to data noise. Table~\ref{tab:metrics_single_gauss} shows the mean and standard deviation of both the reduced $\chi^2$ and the NRMSE, for the different ``ground truth'' input DEM functions used and for each of the six reconstruction methods studied. To emphasize the different types of information that we derive from the mean and the standard deviation of the NRMSE metric, we report in Table~\ref{tab:metrics_single_gauss} the mean and standard deviation in two different columns.

We now discuss the features of the various reconstructions.

\begin{enumerate}

\item The basis pursuit (BP) approach of \cite{2015ApJ...807..143C} (first row of Fig.~\ref{fig:method-comparison-single-gaussian-linear}) is very robust, as evident from the near-congruence of most of the gray curves for the 25 different data realizations and the very low standard deviation values of the NRMSE metric in Table~\ref{tab:metrics_single_gauss}. (The only exception is for the simulation centered on $\log_{10} T_c = 6.3$, for which one of the 25 reconstructions is corrupted by spurious features.) Compared to the other inversion methods, BP appears to be the most accurate in reconstructing the ``ground truth'' $\xi(T)$ profiles with $\log_{10} T_c = 6.1$ and $6.2$, as is evident from the lowest mean NRMSE values (Table~\ref{tab:metrics_single_gauss}). This high degree of accuracy is likely due to the strong similarity between the shapes of the simulated ``ground truth'' configurations and the (Gaussian) basis functions used by this method. However, the fidelity of the reconstructed $\xi(T)$ forms becomes poorer for higher centroid temperatures (third and [especially] fourth columns of Fig.~\ref{fig:method-comparison-single-gaussian-linear}). We note that the mean $\chi^2$ values of the BP reconstructions are systematically larger than those associated with the FIR, SITES, ML, and RML methods. This is probably because the BP method slightly underestimates the peak DEM value and, in order to limit the number of basis functions that are used, it does not compensate for this by adding spurious features at high temperatures, as do SITES and ML (see below).

\item The fast iterative regularized (FIR) methodology of \cite{2013ApJ...771....2P} (second row of Fig.~\ref{fig:method-comparison-single-gaussian-linear}) 
has good fidelity, with a reduced $\chi^2$ of order unity (Table~\ref{tab:metrics_single_gauss}). However, it significantly overestimates the peak in $\xi(T)$ for the low centroid temperature cases, while underestimating it (and shifting the peak to lower temperatures) for cases with higher centroid temperatures. This worsens the agreement with the corresponding ``ground truth'' configurations with respect to BP, as evidenced by the much higher values of the mean NRMSE value (Table~\ref{tab:metrics_single_gauss}). The method is very robust: the NRMSE standard deviations have values that are among the lowest of all the methods studied.

\item The iterative SITES method of \cite{2019SoPh..294..135M} (third row of Fig.~\ref{fig:method-comparison-single-gaussian-linear}) produces quite acceptable reduced $\chi^2$ and NRMSE metrics, especially for the ``ground truth'' profiles centered at $\log_{10} T_c = 6.3$ and $6.4$. However, for low-centroid-temperatures, it recovers a $\xi(T)$ profile that is significantly broader (with a commensurately lower peak value) than the ``ground truth'' profile. This is to be expected: the essence of the SITES algorithm is to add, at each iterative step, a correction to the $\xi(T)$ profile that is based on the difference between the set of forward-fit spectral line intensities and those observed. Any such correction must necessarily introduce structure in $\xi(T)$ that was not present in the previous iteration, and thus, as the number of iterations is increased to provide a better match to the data, the method produces $\xi(T)$ profiles that are progressively broader, leading to a final result that is substantially broader than the ``ground truth.'' Indeed, \cite{2019SoPh..294..135M} note that ``SITES performs poorly for narrow DEM profiles at all temperatures,'' such that ``narrow peaks are found by SITES, but are smoother,'' and this characteristic is evident in the application to AIA data in Section~\ref{sec:AIA-data-application}. The method also introduces a spurious component at very high temperatures ($\log_{10} T \gtrsim 7.2$) in the first three cases, presumably to compensate for the relative inability of the low-temperature component, with its substantial deviation from the ``ground truth'' profile, to adequately fit the data by itself.

\item The regularized approach of \cite{2012A&A...539A.146H} (REG; fourth row of Fig.~\ref{fig:method-comparison-single-gaussian-linear}) typically underestimates the value of the peak and/or shifts it to lower temperatures. The mean reduced $\chi^2$ values tend to be quite large since, because of the underestimated value of the peak DEM, the intensities in the 171~\AA, 193~\AA~and 211~\AA~channels are poorly fitted. However, the reconstructed $\xi(T)$ profiles are generally consistent with the respective ground truth configurations, as evidenced by mean NRMSE values similar to (or even better than) those corresponding to the reconstructions produced by the other methods. The relatively poor data--fitting performance of the method is likely due to the significant emphasis that the penalty term in the minimization equation~(\ref{eq:regularized-equation}) places on smoothness of the recovered $\xi(T)$ profile, even at the expense of loss of fidelity in matching the input data.

\item The (unregularized) maximum likelihood (ML) method based on Eq.~\eqref{eq:ml-solve} (fifth row of Fig.~\ref{fig:method-comparison-single-gaussian-linear}) produces acceptable values of both the mean reduced $\chi^2$ and NRMSE metrics; however, it underestimates the peak $\xi(T)$ for the two lowest centroid temperature cases. This method performs much better for cases with a higher centroid temperature, and in all cases it well reproduces the value of that centroid temperature. However, it systematically creates a spurious artifact at temperatures $\log_{10} T \gtrsim 7.1$. Because of the presence of artifacts at high temperatures (particularly for the two configurations with lowest centroid temperatures), the total Emission Measure $\text{EM} \!\! = \!\! \int \xi(T) \, dT$ of the reconstructions provided by ML and SITES  is \(\sim\)$3$~ to $4$~times larger than the ground truth value of $1 \times 10^{28}$~cm$^{-5}$. The method is also less robust than BP and FIR, with  significant differences between the reconstructed $\xi(T)$ profiles for the $25$~different data realizations and hence higher standard deviation values of the NRMSE metric in Table~\ref{tab:metrics_single_gauss}.

\item The regularized maximum likelihood (RML) method (final row of Fig.~\ref{fig:method-comparison-single-gaussian-linear}) produces reduced $\chi^2$ values that are all of order unity, indicating that it is able to fit the data with high fidelity. It also reproduces the peak intensities and centroid temperatures of the input $\xi(T)$ with $\log_{10} T_c = 6.1$~and~$6.2$ (first two columns of Fig.~\ref{fig:method-comparison-single-gaussian-linear}) with greater accuracy compared to FIR, SITES and ML, as indicated by the lower mean NRMSE values (Table~\ref{tab:metrics_single_gauss}). However, the RML method does progressively underestimate the high-temperature wings (at $\log_{10} T \simeq 6.5$) in the third column, and significantly underestimates the $\log_{10} T = 6.4$ centroid temperature in the last column. Indeed, underestimation of $\xi(T)$ at such temperatures is a common element of all of the reconstruction methods tested (except ML and, to a lesser extent, SITES). Reference to Fig.~\ref{fig:aia_t_resp} reveals why this may be the case: temperatures $\log_{10} T \simeq 6.3 - 6.7$ correspond to rather low values of the temperature response curves $K_i(T)$ in all channels, so that fitting the data in any observed channel is more straightforwardly accomplished by adding a relatively small amount of emission measure at other temperatures. Finally, RML is a very robust method, as evidenced by the values of the NRMSE standard deviation, which, except for the configuration centered at $\log_{10} T_c = 6.4$, are among the lowest produced by any method.

\end{enumerate}

\begin{figure*}[pht]
\centering
\includegraphics[height=\textwidth, angle=90]{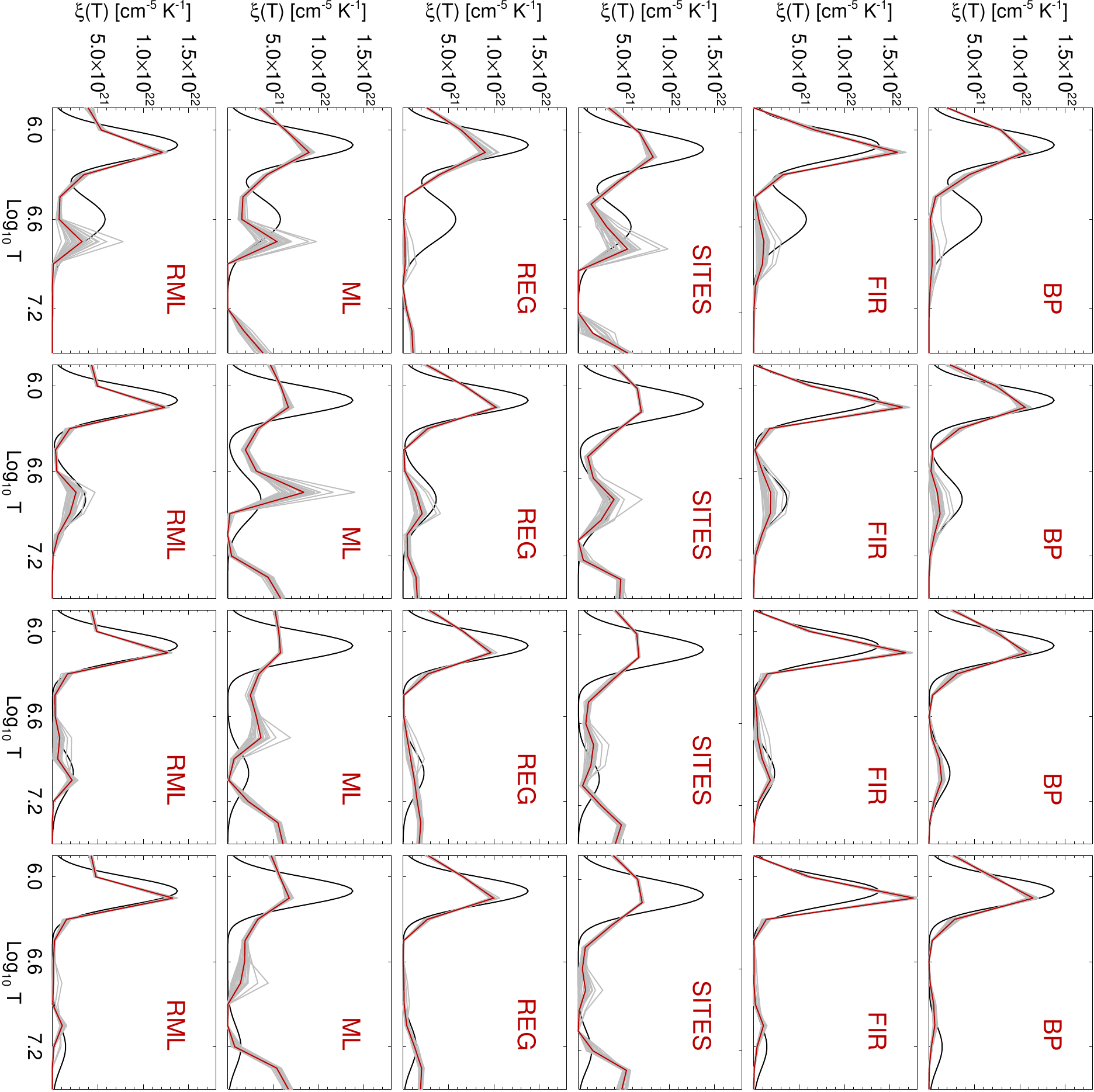}
\caption{Results of the double-Gaussian simulation tests. \textit{From top to bottom:} Reconstructions of a simulated double Gaussian $\xi(T)$ profile by means of the basis pursuit (BP) technique \citep{2015ApJ...807..143C}, the FIR inversion method of \cite{2013ApJ...771....2P}, the iterative SITES technique \citep{2019SoPh..294..135M}, the regularization technique of \cite{2012A&A...539A.146H} (REG), the ML method defined in Eq.~\eqref{eq:ml-solve}, and finally our proposed RML technique. The black solid lines represent the ``ground truth'' configurations, which consist of double Gaussian profiles. The left Gaussian has a total Emission Measure $\text{EM} = \int \xi(T) \, dT = 10^{28}$~cm$^{-5}$ and a standard deviation of $0.1$~dex, and is centered on $\log_{10} T_c = 6.1$, while the right Gaussian has a total Emission Measure $\text{EM} = \int \xi(T) \, dT = 2 \times 10^{28}$~cm$^{-5}$ and a standard deviation of $0.15$~ dex, and is centered on (from left to right) $\log_{10} T_c = 6.6, 6.8, 7.0$ and~$7.2$. The gray lines represent reconstructions from $25$~different realizations of Poisson noise affecting the data; the red lines represent the mean values of these $25$~reconstructions. The $\xi(T)$ profiles are plotted as a function of the (base~$10$) logarithm of the temperature~(K).}
\label{fig:method-comparison-double-gaussian-linear}
\end{figure*}

\subsubsection{Double Gaussian forms}\label{sec:double_gaussian}

Fig.~\ref{fig:method-comparison-double-gaussian-linear} shows the results obtained with simulated ``ground truth'' $\xi(T)$ that take the form of two Gaussian functions of $\log_{10} T$. In application of the methods to real data (see Sect.~\ref{sec:AIA-data-application} below), it is apparent that ascertaining the veracity of recovered high-temperature ($\log_{10} T_c \!\! \gtrsim \!\! 7.0$) components in $\xi(T)$ is important. Hence, in the simulation test shown in Fig.~\ref{fig:method-comparison-double-gaussian-linear}, the lower-temperature Gaussian has a total Emission Measure $\text{EM} \! = \! \int \xi(T) \, dT \!\! = \!\! 10^{28}$~cm$^{-5}$ and a standard deviation of $0.1$~dex, and is centered on $\log_{10} T_c \! = \! 6.1$, while the higher-temperature Gaussian has a total Emission Measure $\text{EM} \!\! = \!\! \int \xi(T) \, dT \!\! = \!\! = 2 \times 10^{28}$~cm$^{-5}$ and a standard deviation of $0.15$~ dex, and is centered on four different values of $\log_{10} T_c = 6.6, 6.8, 7.0$ and~$7.2$. Again, the gray lines represent the reconstructions from $25$~different data realizations, and the red lines represent the mean values of these $25$~reconstructions. Table~\ref{tab:metrics_double_gauss} shows the means and standard deviations of the same two validation metrics used in Table~\ref{tab:metrics_single_gauss}.

\begin{table*}[ht]
\centering
\resizebox{\linewidth}{!}{\begin{tabular}{ccccccccccccc}
\toprule
Method &\multicolumn{3}{c}{$\log_{10} T_c=6.6$} & \multicolumn{3}{c}{$\log_{10} T_c=6.8$} & \multicolumn{3}{c}{$\log_{10} T_c=7.0$} & \multicolumn{3}{c}{$\log_{10} T_c=7.2$}\\
\midrule
&Red. $\chi^2$ &\begin{tabular}{@{}c@{}}NRMSE \\ Mean\end{tabular} &\begin{tabular}{@{}c@{}}NRMSE \\ Std. Dev.\end{tabular} &Red. $\chi^2$ &\begin{tabular}{@{}c@{}}NRMSE \\ Mean\end{tabular} &\begin{tabular}{@{}c@{}}NRMSE \\ Std. Dev.\end{tabular}  &Red. $\chi^2$ &\begin{tabular}{@{}c@{}}NRMSE \\ Mean\end{tabular} &\begin{tabular}{@{}c@{}}NRMSE \\ Std. Dev.\end{tabular} &Red. $\chi^2$ &\begin{tabular}{@{}c@{}}NRMSE \\ Mean\end{tabular} &\begin{tabular}{@{}c@{}}NRMSE \\ Std. Dev.\end{tabular}\\
\cmidrule{2-13}
BP &9.1 $\pm$ 1.3 &0.69 &0.03 &12 $\pm$ 0.9 &0.47 &0.06 &11 $\pm$ 0.7 &0.38 &0.02 &10 $\pm$ 0.4 &0.41 &0.01 \\
FIR &2.0 $\pm$ 0.8 &0.73 &0.05 &2.3 $\pm$ 0.7 &0.60 &0.08 &2.5 $\pm$ 0.7 &0.68 &0.07 &2.2 $\pm$ 0.8 &0.84 &0.07 \\
SITES &3.5 $\pm$ 1.1 &0.84 &0.05 &1.7 $\pm$ 0.3 &1.04 &0.03 &1.0 $\pm$ 0.1 &1.19 &0.02 &1.3 $\pm$ 0.6 &1.23 &0.03 \\
REG &33 $\pm$ 18 &0.80 &0.03 &1.3 $\pm$ 0.2 &0.50 &0.04 &1.2 $\pm$ 0.1 &0.58 &0.03 &1.5 $\pm$ 0.7 &0.54 &0.02 \\
ML &2.5 $\pm$ 1.0 &0.79 &0.06 &0.5 $\pm$ 0.1 &1.36 &0.09 &0.17 $\pm$ 0.04 &1.60 &0.04 &0.3 $\pm$ 0.1 &1.50 &0.04 \\
RML &0.8 $\pm$ 0.1 &0.69 &0.03 &0.8 $\pm$ 0.1 &0.58 &0.02 &0.7 $\pm$ 0.1 &0.66 &0.01 &0.7 $\pm$ 0.1 &0.73 &0.01 \\
\bottomrule
\end{tabular}}
\caption{As for Table~\ref{tab:metrics_single_gauss}, but for the double Gaussian ``ground truth'' model tests. The columns are labeled by the (logarithm of the) centroid temperature of the high-temperature Gaussian component; all models also have a lower-temperature Gaussian component with $\log_{10} T_c = 6.1$ (see Sect.~\ref{sec:double_gaussian} and Figure~\ref{fig:method-comparison-double-gaussian-linear}).}\label{tab:metrics_double_gauss}
\end{table*}

\begin{enumerate}

\item The basis pursuit (BP) approach of \cite{2015ApJ...807..143C} generally underestimates the intensity of both the low-temperature and high-temperature components, the latter being especially obvious in the first column, corresponding to a high-temperature component with the lowest centroid temperature of the four cases considered. The resulting mean NRMSE values (Table~\ref{tab:metrics_double_gauss}) are hence considerably larger than those for the first two cases presented in the single-Gaussian experiment. This underestimation of the peak intensity of the low temperature component is likely the reason for the rather large mean $\chi^2$ values (Table~\ref{tab:metrics_double_gauss}). With the exception of the test presented in the first column, and, to a lesser extent, the test presented in the fourth column, the BP method does correctly identify the approximate centroid temperatures of both low- and high-temperature components. As measured by the standard deviation of the NRMSE values, the robustness of the method for the double-Gaussian case is comparable to that for the single-Gaussian case. However, as expected from the more complex configuration that has to be reconstructed, the NRMSE standard deviation values corresponding to the first two configurations presented in Table~\ref{tab:metrics_double_gauss} are slightly larger than most of those obtained in the single-Gaussian test.

\item The FIR approach \citep{2013ApJ...771....2P} typically underestimates the high-temperature component, but apparently compensates for this by significantly overestimating the intensity of the low-temperature component. Similar to the results for the single-Gaussian experiment, the method is able to fit the data with high fidelity, as shown by mean reduced $\chi^2$ values that are generally of order unity. It also reproduces the ``ground truth'' profile rather accurately, as evident from mean NRMSE values (Table~\ref{tab:metrics_double_gauss}) that are among the lowest of all the methods considered. With regard to robustness, there is considerable variation in the reconstructions for different data realizations, particularly with reference to the high-temperature component. This results in a standard deviation of the NRMSE metric which is almost always larger than those for the other methods.

\item In all cases the iterative SITES method of \cite{2019SoPh..294..135M} significantly broadens the low-temperature Gaussian component and concomitantly underestimates its peak value, similar to the method's performance in the single-Gaussian tests (Figure~\ref{fig:method-comparison-single-gaussian-linear}). To compensate for this, and still reproduce reasonably low reduced $\chi^2$ values (see Table~\ref{tab:metrics_double_gauss}), the method introduces a spurious feature at $\log_{10} T \gtrsim 7.2$. This behavior, similar to that of the ML method (see below), results in mean NRMSE values that are generally quite large (near unity; see Table \ref{tab:metrics_double_gauss}). Similar behavior is found in Figure~5 of \cite{2019SoPh..294..135M}, who note that ``the hot peak is well-fitted by SITES, but the fit for the cool peak is poor.'' For the case with a high-temperature component at $\log_{10} T_c \!=\! 6.8$ (second column of Figure~\ref{fig:method-comparison-double-gaussian-linear}) SITES does a reasonable job of fitting this high-temperature component, but its performance is not as good for other centroid temperatures for the high-temperature component.

\item The regularized inversion technique (REG) of \cite{2012A&A...539A.146H} significantly underestimates both low- and high-temperature components, especially for lower values of the centroid temperature of the high-temperature component. This is particularly evident in the case presented in the first column of Fig.~\ref{fig:method-comparison-double-gaussian-linear}, where the method consistently produces a rather large $\chi^2$ value. Interestingly, in all the other double-Gaussian cases, REG systematically fits the data with remarkable fidelity, with mean reduced $\chi^2$ values very close to unity. The method is also among the most accurate and robust, as evident from the low means and standard deviations of the NRMSE metric. Consistently, the $\xi(T)$ profiles for the 25 different data realizations are very similar, even to the point of consistently reconstructing the spurious high-temperature features at $\log_{10} T \gtrsim 7.2$. Similar artifacts are also present in the reconstructions shown in the first two columns of Fig.~\ref{fig:method-comparison-single-gaussian-linear}, although, in these cases, they are less prominent.

\item The (unregularized) maximum likelihood (ML) approach significantly underestimates the magnitude of the low-temperature component, and severely overestimates both the magnitude and width of the high-temperature component, in all cases creating features at temperatures well in excess of any that are actually present in the ``ground truth'' $\xi(T)$ profile. The presence of these artifacts is reflected in the large (near unity) mean values for the NRMSE metric. Similar to the SITES reconstructions, these spurious features are most likely introduced to compensate for the underestimated peak of the low-temperature component, as evidenced by the low mean reduced $\chi^2$ value (see Table~\ref{tab:metrics_double_gauss}). Interestingly, similar to REG, ML deduces these (erroneous) high-temperature $\xi(T)$ features rather robustly, as evidenced by values of the NRMSE standard deviation that are comparable to those provided by the other methods. We finally note that, due to the presence of high temperature artifacts, the total Emission Measure of the reconstructions provided by ML and SITES ranges from $8 \times 10^{28}$~cm$^{-5}$ to $1.5 \times 10^{29}$~cm$^{-5}$, and from $1.0 \times 10^{29}$~cm$^{-5}$ to $1.2 \times 10^{29}$~cm$^{-5}$, respectively; all these are an order of magnitude larger than the ground truth value of $3 \times 10^{28}$~cm$^{-5}$.

\item The regularized maximum likelihood approach (RML) tends to slightly overestimate the centroid temperature of the low--temperature component. However, together with SITES, it is the most accurate in reconstructing both the centroid and peak value of the high temperature components for the tests presented in the second column of Figure \ref{fig:method-comparison-double-gaussian-linear}. Further, only RML, SITES and ML suggest the presence of the (actual) high-temperature component in the test case considered in the first column, albeit without always accurately reproducing the centroid temperature of this component. A comparison with the results obtained by means of the (un--regularized) ML method shows the effectiveness of the adopted regularization, which suppresses spurious high-temperature features present in the ML reconstructions. Table~\ref{tab:metrics_double_gauss} shows that RML achieves a good compromise between fidelity to the data and accuracy (similarity to the ground truth configuration), with mean reduced $\chi^2$ values that are systematically close to unity and mean NRMSE values that are systematically among the lowest for the methods considered. The RML method also proves to be very robust, with NRMSE standard deviations that are systematically the lowest (Table~\ref{tab:metrics_double_gauss}).  Most importantly, comparing the single-Gaussian results of Fig.~\ref{fig:method-comparison-single-gaussian-linear} and the double-Gaussian results of Fig.~\ref{fig:method-comparison-double-gaussian-linear}, it is apparent that the RML method recovers a (suitably scaled and positioned) high-temperature component if and only if one is actually present.

\end{enumerate}

\begin{figure*}[ht]
\centering
\includegraphics[height=\textwidth, angle=90]{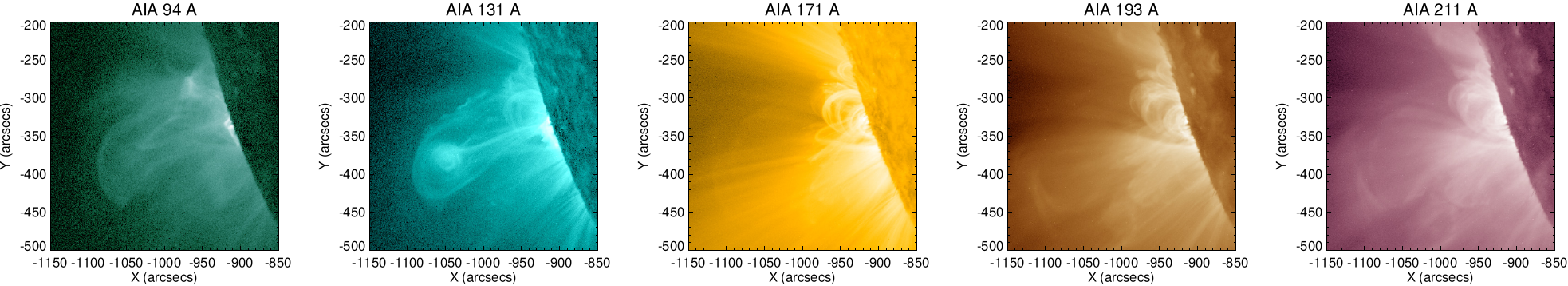}
\caption{Images recorded by AIA on November~3, 2010 around 12:15:02~UT in the 94 \AA, 131 \AA, 171 \AA, 193 \AA, and 211 \AA\ channels, from left to right, respectively.}
\label{fig:aia_maps}
\end{figure*}

\begin{figure*}[h!]
\centering
\includegraphics[width=0.69\textwidth]{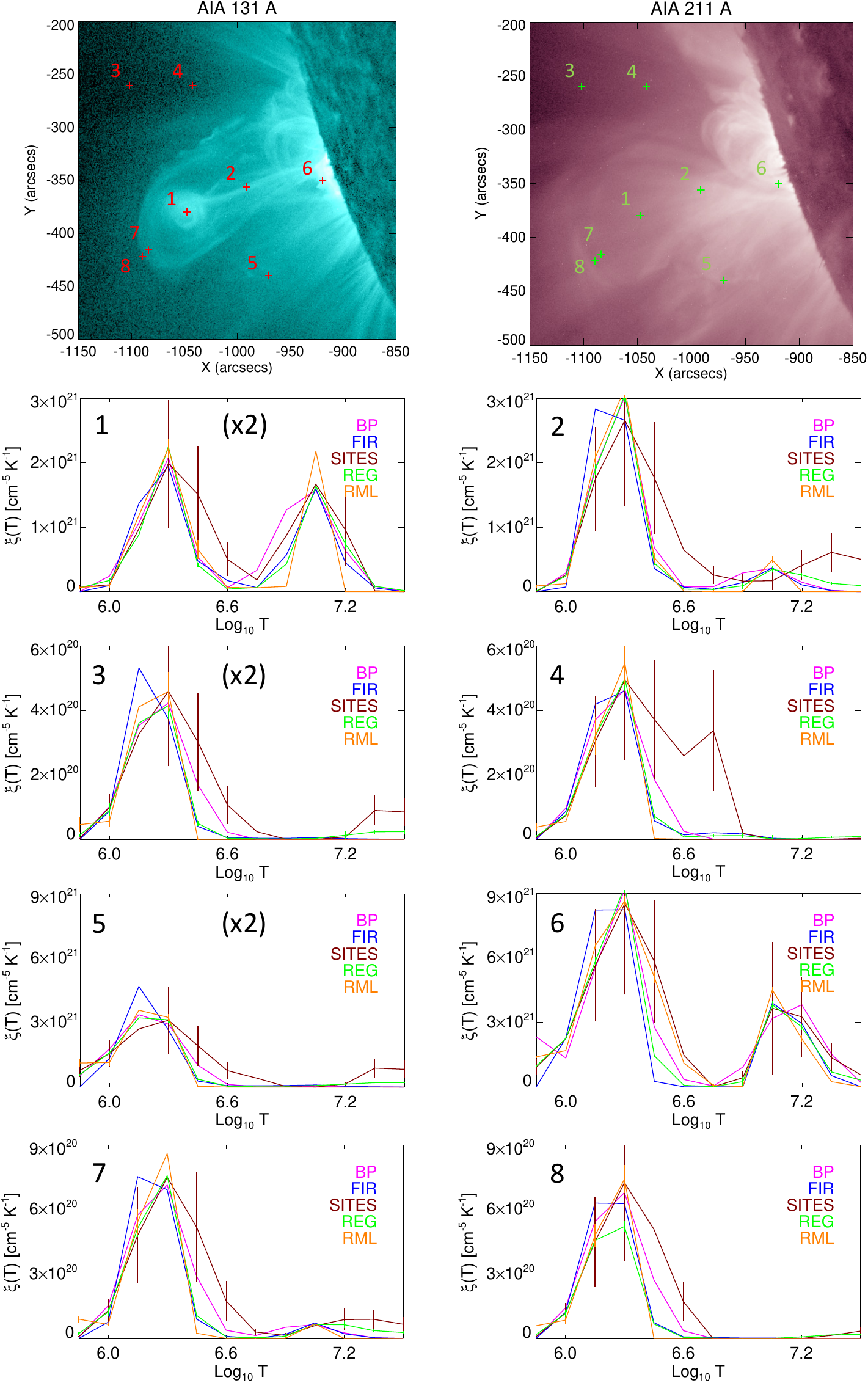}
\caption{DEM profiles reconstructed from observed AIA count rates in selected pixels of the 2010 November~3 event. \textit{Top row:} AIA images recorded around 12:15:02~UT in the 131~\AA~and 211~\AA~channels (left and right panel, respectively). The numbered crosses denote the location of the pixels selected for a comparison of the $\xi(T)$~profiles reconstructed by the different methods. \textit{Second to fifth row:} $\xi(T)$~profiles associated with Pixels~1 through~8. In each panel, the profiles reconstructed by the BP, FIR, SITES, REG and~RML methods are plotted in magenta, blue, brown, green, and orange, respectively. The $\pm 1 \sigma$ uncertainties associated with the SITES, REG and~RML reconstructions have been added as error bars at each temperature point in the pertinent reconstruction. The intensities of the reconstructions of Pixels~1, 3 and~5 have been multiplied by a factor of~$2$. The intensity axes are linear, while the temperature axes are in terms of $\log_{10} T$~(K).}
\label{fig:aia_dems}
\end{figure*}

\begin{table*}[ht]
\centering
\begin{tabular}{rcccccccc}
\toprule
Method  &Pixel 1 &Pixel 2 &Pixel 3 &Pixel 4 &Pixel 5 &Pixel 6 &Pixel 7 &Pixel 8 \\
\midrule
BP &2.13 &1.96 &1.05 &1.62 &2.13 &2.45 &1.49 &1.44\\
FIR  &1.11 &2.82 &0.57 &0.94 &3.98 &0.67 &1.11 &0.83\\
SITES &0.40 &3.00 &0.08 &0.14 &0.31 &1.22 &0.14 &0.08\\
REG &1.03 &1.27 &1.15 &1.18 &0.90 &2.35 &1.16 &11.36\\
RML &0.55 &0.65 &0.35 &0.72 &0.99 &0.70 &0.86 &0.94\\
\bottomrule
\end{tabular}
\caption{Reduced $\chi^2$ values for the various pixels in the 2010~November~3 solar eruptive event (Fig.~\ref{fig:aia_dems}), for five different DEM reconstruction algorithms.}
\label{tab:chi2}
\end{table*}

\begin{figure*}[ht]
\centering
\includegraphics[width=\textwidth]{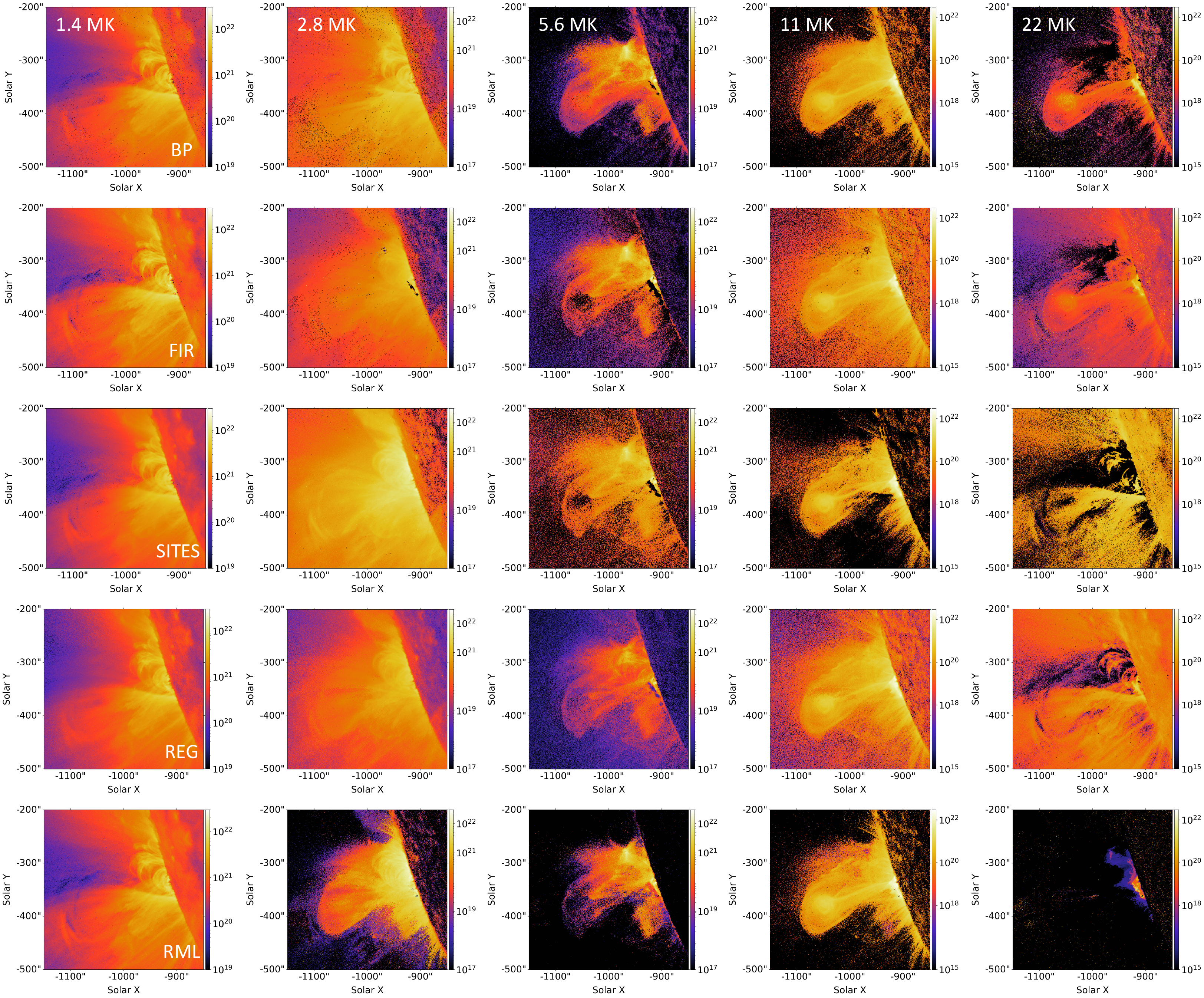}
\caption{DEM maps reconstructed from the AIA images shown in Fig.~\ref{fig:aia_maps}.
\textit{Rows, from top to bottom:} the BP method \citep{2015ApJ...807..143C}, the FIR method \citep{2013ApJ...771....2P}, the SITES method \citep{2019SoPh..294..135M}, the REG method \citep{2012A&A...539A.146H}, and our proposed RML method. The images of each column correspond to the same temperature, which is reported in the top--left corner of the first row panels. The same color map is shared by each of the images in the same temperature column, with different (logarithmic) scalings for the different temperatures.
}
\label{fig:dem_bp}
\end{figure*}

\subsection{Application to AIA data}\label{sec:AIA-data-application}

Here we compare\footnote{Small differences between the results presented in this paper and those in \cite{2013A&A...553A..10H} result from the fact that we do not consider the AIA 335~\AA\ channel in reconstructing the $\xi(T)$ profiles. Although, as discussed in Sect.~\ref{section:simulated_data} above, the 335~\AA\ channel has a small emissivity function over the temperature range of interest (Fig.~\ref{fig:aia_t_resp}) and hence can lead to the introduction of spurious features in the recovered $\xi(T)$ profiles, this very consideration means that its influence on the recovered $\xi(T)$ profiles is not entirely negligible.} the $\xi(T)$ reconstructions for a selected set of pixels in the 2010~November~3 12:15~UT event previously studied by \cite{2013A&A...553A..10H}. Fig.~\ref{fig:aia_maps} shows the general morphology of the region, which produced both a flare and an erupting flux rope, at 12:15:02~UT in several SDO/AIA channels. Eight pixels (see Fig.~\ref{fig:aia_dems}) were selected by \cite{2013A&A...553A..10H} as representative of different features in the field of view. In Fig.~\ref{fig:aia_dems} we show the $\xi(T)$ reconstructions for each of these pixels by the various reconstruction methods (excluding the problematic unregularized ML method) considered above. Below we discuss and compare the general features of the results obtained.

\begin{itemize}

    \item \textit{Pixel 1 -- Core of the erupting plasmoid:} The various reconstructions are quite similar, and clearly show two components to the emission, a relatively cool one at around $1.5 \times 10^6$~K and a hotter one at around $10^7$~K. We agree with the interpretation of \cite{2013A&A...553A..10H} that these components probably represent the background corona along the line of sight, and the plasmoid emission, respectively. The methods provide consistent reconstructions of the low--temperature component, although SITES returns a significantly broader high temperature wing. For the high--temperature component, the centroid is consistently retrieved by the different techniques, while its reconstructed peak value, width and skewness vary somewhat from method to method.
   
    \item \textit{Pixel 2 -- Filament/stem behind the plasmoid:} Here the relatively cool background line-of-sight emission is about $3 \times $ more intense than for Pixel~1, while the hot component is about $3 \times$~less intense. Although most of the methods agree on the temperature of the two components, the FIR method appears to broaden the low-temperature component and shift its centroid temperature downward slightly, while, as expected from the results of Sections~\ref{sec:single_gaussian} and~\ref{sec:double_gaussian}, the SITES method produces a much broader profile with a lower value of $\xi(T)$ at the centroid.
    
    \item \textit{Pixel 3 -- High corona away from the event:} This pixel contains a relatively small amount of emission (peak value of $\xi(T) \simeq 2 \times 10^{20}$~cm$^{-5}$~K$^{-1}$) at a temperature around $\log_{10} T = 6.2$, similar to the temperature of the background corona components in Pixels~1 and~2. The BP, REG and RML methods agree on the centroid temperature of this component, although BP produces an enhanced high--temperature wing and RML returns a higher estimate of the peak value of $\xi(T)$. However, the reconstructions by BP and REG both fall within the $\pm 1 \sigma$ error bars associated with the RML reconstruction, suggesting that this is simply due to statistical uncertainty in the data. The FIR and SITES methods again (cf. Pixels~1 and~2) produce a broader peak with an enhanced low (high) temperature wing compared to the other methods.
    
    \item \textit{Pixel 4 -- Corona away from the event:} All methods agree very well as to the presence and intensity of a relatively cool $\log_{10} T \simeq 6.2$ component; however, the SITES method again produces a much broader profile, and considerations similar to those for Pixel 3 apply to the enhanced high--temperature wing produced by BP and the higher estimate of the peak flux returned by RML.
      
    \item \textit{Pixel 5 -- Corona near the event:} Again, all methods agree very well as to the presence and intensity of a relatively cool $\log_{10} T \simeq 6.2$ K component; however, the FIR method substantially overestimates the low-temperature ``wing'' of this component relative to the other three methods, while the SITES method overestimates the high-temperature wing. Further, similar to Pixels~3 and~4, the BP method reconstructs a $\xi(T)$ profile that is more skewed toward higher temperature values.
    
    \item \textit{Pixel 6 -- Low corona flare emission:} All methods agree very well with regard to both the relatively cool ($\log_{10} T \simeq 6.2$) and hot ($\log_{10} T \simeq 7.1$) components present. Again, the FIR method creates a low-temperature $\xi(T)$ component that is broader (and skewed toward lower temperatures) than those of the other methods. Further, both the BP and the SITES reconstructions of the high--temperature component show an enhancement in the high--temperature wing, while the other methods agree quite well in terms of the centroid value, peak intensity and width of the reconstructed high--temperature component.
    
    \item \textit{Pixel 7 -- Envelope just ahead of the plasmoid:} Here the FIR method once again produces a centroid of the low--temperature component that is broader and slightly shifted toward lower temperatures compared to the other reconstructions, while the SITES method again produces a broader component with an enhanced high-temperature wing. Also, the BP method retrieves an enhanced high-temperature ``wing'' for this component. There is also evidence of a weak high-temperature ($\log_{10} T \simeq 7.0$) component. Given the location of this material, it is not unreasonable to expect additional heating there, given, for instance, the findings of \cite{2020FrASS...7....1M} that ``the CME is in the heat-releasing state (i.e., entropy loss) throughout its journey from the Sun to Earth.''
    
    \item \textit{Pixel 8 -- Further ahead of the plasmoid:} Here all methods agree that there is a low-temperature component with an intensity similar to that of Pixel~7. The BP and SITES methods both  show an enhancement in the high--temperature wing of the $\xi(T)$ profile. The 131 \AA\ image shows that this pixel is at the leading edge of the erupting material, providing (similar to Pixel~7) a plausible explanation for such enhanced heating of the low-temperature component. 

\end{itemize}

In any ill-posed inversion problem, there is a necessary tradeoff between fidelity to the data and accuracy in the recovered solution. The fidelity to the AIA data is determined by the reduced $\chi^2$ values shown in Table~\ref{tab:chi2}, which are based on a comparison of the original data with the line intensities obtained by substituting the recovered $\xi(T)$ profile into Eq.~\eqref{eq:DEM-def}. Most of the $\chi^2$ values for the eight pixels are of order unity, showing that the data is well-fit but not overfitted at the expense of the plausibility of the $\xi(T)$ profile. Notable exceptions are many of the SITES reconstructions, which have reduced $\chi^2$ values that are generally below unity: the resulting overfitting of noisy data could be among the reasons behind the presence of spurious artifacts in the reconstructions. Other notable exceptions occur in application of the FIR method to Pixel~5, where the method creates a more intense peak compared to the reconstructions by the other methods, and in application of the REG method to Pixel~8, where the reconstruction has a significantly lower peak value compared to those of the other methods.

\subsection{Differential emission measure maps}\label{sec:dem-maps}

\begin{figure*}[ht]
\centering
\includegraphics[width=\textwidth]{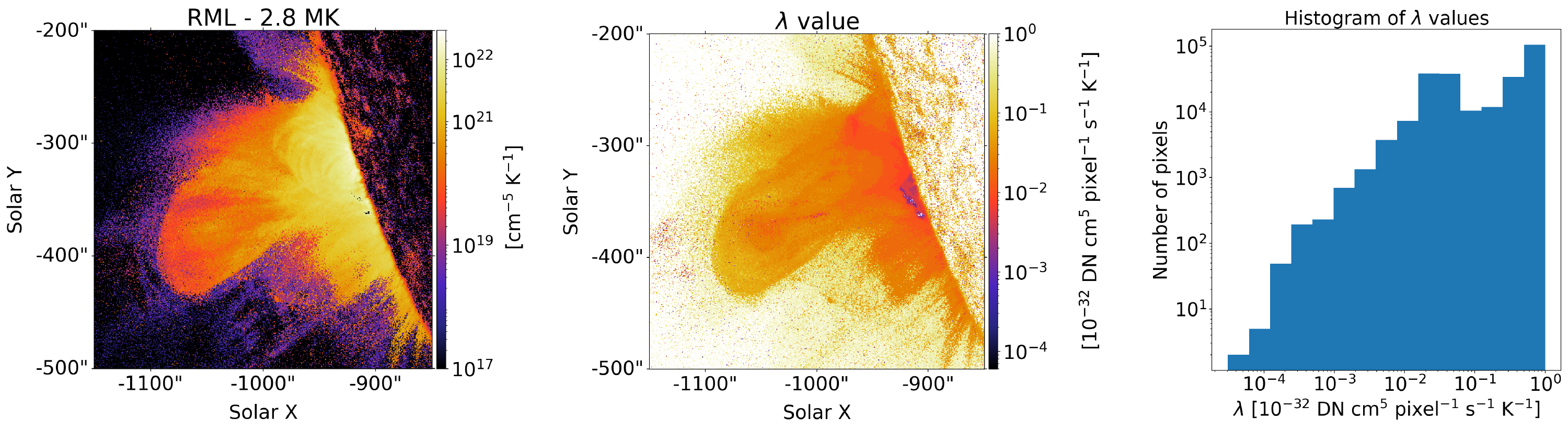}
\caption{Distribution of values of the RML regularization parameter. \textit{Left panel:} Reconstructed DEM map corresponding to $T = 2.8$ MK, produced by the RML method. \textit{Middle panel:} Value of the regularization parameter $\lambda$ used at each pixel within the image. \textit{Right panel:} Histogram of the $\lambda$ values selected by RML for the \(\sim\)250,000 $0.6^{{\prime}{\prime}}$ $\times$ $0.6^{{\prime}{\prime}}$ pixels within the image.
}
\label{fig:lambda_image}
\end{figure*}

In Fig.~\ref{fig:dem_bp} we show the DEM maps reconstructed by each of the five reconstruction methods considered (rows), at five different temperatures (columns), logarithmically scaled with a factor of two between successive temperatures. In each map, the pixel value corresponds to the the value of the DEM profile $\xi(T)$ calculated for that specific pixel at the temperature in question. As pointed out by \cite{2013A&A...553A..10H}, for some temperatures the DEM maps closely resemble the images from one of the AIA channels: for example, the $T = 1.4$~MK map closely resembles the 171~\AA\ image, and the $T = 11$~MK map closely resembles the 131~\AA\ image. These close matches reflect the well-defined peaks in the response curves for those channels at those respective temperatures (Fig.~\ref{fig:aia_t_resp}). However, the DEM maps for other temperatures reveal features that are not as apparent in the individual SDO/AIA images, such as the large region of emission extending out to coordinates $(x \simeq - 1100, y \simeq [-450 : -350])$ above the eastern limb in the $T = 2.8$~MK maps. This feature is reproduced consistently by all the reconstruction methods (see Fig.~\ref{fig:dem_bp}), confirming its reality and highlighting the generally complicated relationship between temperature and emitted wavelength, particularly for SDO/AIA channels with a relatively broad temperature response. We also note (cf. remarks in Sect.~\ref{sec:intro}) that the DEM~maps produced by the BP method (see Fig.~\ref{fig:dem_bp}) contain $253$ ``null'' pixels, showing points where the simplex optimization method adopted by BP has not been able to find a solution that satisfies all the required constraints. Finally, looking at the high temperature (22~MK) maps, we see that the SITES method produces a relatively high amount of emission at these temperatures (cf. the results of Sects.~\ref{sec:single_gaussian} and~\ref{sec:double_gaussian}). On the other hand, the RML method, consistent with the way it is designed (with a regularization term that acts to suppress the presence of $\xi(T)$ features at high temperatures; see remarks following Eq.~\eqref{eq:rml}), produces substantially less emission at such high temperatures.

In Fig.~\ref{fig:lambda_image} we show the RML 2.8~MK DEM map, together with a map of the regularization parameter $\lambda$ (Eqs.~\eqref{eq:rml} and~\eqref{eq:rml_solve}) used at each pixel within the image and a histogram of the $\lambda$ values used throughout the image. In general, bright pixels have a higher signal-to-noise ratio and hence require a lower degree of regularization during the inversion process, and Fig.~\ref{fig:lambda_image} indeed shows that the more intense regions of the flare are associated with smaller values of the regularization parameter $\lambda$. The number of pixels that use a given value of $\lambda$ has an approximately monotonic dependence on $\lambda$: most pixels require a high degree of regularization to produce a physically acceptable result, while a few pixels in the most intense regions of the flare produce a satisfactory $\xi(T)$ profile with little to no regularization required. It is important to note, however, that the value of the regularization parameter is not simply determined by the statistical quality of the data: the optimal regularization parameter value selected by the Morozov discrepancy principle also depends on the ``shape'' of the DEM profile to be reconstructed. Finally, we note that, given the several orders of magnitude spanned by the values of the regularization parameter $\lambda$ necessary to yield  acceptable reduced $\chi^2$ values in all pixels, approximating the regularization parameter with a constant value would result in a substantial decrease of the RML performance, both in terms of fidelity to the data and accuracy of the reconstructed $\xi(T)$ profiles.

\section{Summary}\label{sec:summary}

The results of Sects.~\ref{section:simulated_data} and~\ref{sec:AIA-data-application} show that, both for simulated and actual AIA data, the $\xi(T)$ reconstructions obtained using the regularized maximum likelihood (RML) method described in Sect.~\ref{sec:method-description} are broadly compatible with those of other methods; in no case does the RML~method create a $\xi(T)$ profile that is an ``outlier,'' and, with the possible exception of the very weak (and hence statistically uncertain) Pixel~3, in no case is its $\chi^2$ value unacceptably large (corresponding to over-smoothing) or unacceptably small (corresponding to over-fitting of data).

As evidenced by the reconstructed $\xi(T)$ profiles constructed from different (Poisson-noise) realizations of the data (Figs.~\ref{fig:method-comparison-single-gaussian-linear} and \ref{fig:method-comparison-double-gaussian-linear}), and by comparisons of the spectral line data produced by using the ``ground truth'' and recovered $\xi(T)$ profiles in Eq.~\eqref{eq:DEM-def}, the RML~method\footnote{Both IDL and Python codes that implement the RML method are available at \url{https://github.com/paolomassa/WAFFLE.git}.} is characterized by excellent performance in all three areas of concern: fidelity to the data, accuracy in the reconstructed $\xi(T)$ profiles, and robustness in the presence of data noise. Further, it is straightforward to implement and computationally efficient, taking\footnote{When a more efficient rule for the selection of the regularization parameter (which does not involve performing multiple reconstructions as is the case when the Morozov discrepancy principle is employed) is implemented, the computational time for reconstructing $500\times 500$ $\xi(T)$~profiles (without uncertainty estimation) should decrease to less than $10$~s.} about $35$~s to reconstruct $\xi(T)$ profiles for $500\times 500$ pixels ($\sim 7000$ $\xi(T)$~reconstructions per s) on an Apple MacBook Pro M1 (Chip Apple M1, CPU 8-core) processor, without considering need to reconstruct solutions for multiple data realizations in order to estimate the uncertainty on the solution. Further, it does not require an a priori choice of a parametric functional form and, very importantly for this particular application, always generates a nonnegative solution without the need to impose a posteriori adjustments on the solutions obtained \citep[cf.][]{2013ApJ...771....2P}. We conclude that it is an appropriate  method to use in the construction of $\xi(T)$ profiles from pixel-by-pixel AIA data.

In future work we will develop a more general version of RML that is applicable to other data sets, such as those from EIS or XRT, in order to better constrain \citep[cf.][]{2013A&A...553A..10H} the overall solution of the DEM reconstruction problem. Given the general features of the RML~reconstructions (existence, fidelity, accuracy, robustness, and nonnegativity), this DEM reconstruction method is particularly suitable for the application of machine learning tools to solar data. Furthermore, the computational efficiency of the method will allow us to generate, from AIA data in near real-time, four-dimensional data hypercubes $\xi(x,y; T; t)$ \citep[see Sect.~5 of][]{10.3389/fspas.2022.1040099} that represent a time series of DEM maps. Application of machine learning tools to such hypercubes, each labeled with the eventual level of activity (e.g., maximum GOES level, duration of high levels of emission, total energy released) that results in the event represented by the data hypercube in question, can be used to identify both morphological and thermodynamic precursors of solar activity \citep[see, e.g.,][]{2020arXiv201106433G}, thus providing a promising tool for predicting the timing, location, and characteristics of solar eruptive events.

\section*{Acknowledgements}
We thank Lars Hebenstiel for several useful discussions on the features, strengths and weaknesses of various DEM reconstruction algorithms, and Dr. Huw Morgan for several constructive suggestions on improving the manuscript. We also thank the authors of the BP \citep{2015ApJ...807..143C}, FIR \citep{2013ApJ...771....2P}, and SITES \citep{2019SoPh..294..135M} methods for making their codes freely available, thus allowing us to use them in the comparison of various methods. PM and AGE were supported by NASA Kentucky under award number 80NSSC21M0362; IGH and EPK were supported by STFC consolidated grant ST/T000422/1.

\bibliographystyle{aa}
\bibliography{biblio}

\end{document}